\documentclass[aps,twocolumn,superscriptaddress,amsmath,amssymb,pra,floatfix
,10pt]{revtex4-2}
\usepackage{url}
\usepackage{placeins}
\usepackage{enumerate}
\usepackage{multirow}
\usepackage{tabu}
\usepackage{textcomp,booktabs}
\usepackage{colortbl}
\usepackage[T1]{fontenc}
\usepackage{subcaption}

\usepackage[hmargin=.7in,vmargin=.7in]{geometry}
\usepackage{newtxtext}
\usepackage{dcolumn}
\usepackage{bm}

\usepackage{centernot}
\usepackage{mathtools}
\usepackage{stmaryrd}
\usepackage{caption}
\usepackage{ragged2e}
\usepackage{physics}
\usepackage{dsfont}
\usepackage{mathrsfs}
\usepackage{tikz}
\usepackage{pgfplots}
\pgfplotsset{compat=1.7}
\usepackage{makecell}
\usepackage{amsmath}
\usepackage{bbold}

\usepackage{mwe} 

\makeatletter

\newcommand{\mA}{\mathcal{A}}

\newcommand{\mT}{\mathcal{T}}
\newcommand{\mE}{\mathcal{E}}

\newcommand{\mL}{\mathcal{L}}

\newcommand{\1}{\mathbb{1}}

\newcommand{\bal}{\begin{equation}\begin{aligned}}
\newcommand{\eal}{\end{aligned}\end{equation}}

\newcommand*{\rom}[1]{\expandafter\@slowromancap\romannumeral #1@}
\makeatother

\usepackage{amsthm}
\newtheorem*{thm*}{Theorem}

\DeclareMathOperator*{\argmin}{arg\,min}

\usetikzlibrary{arrows}
\usepackage{verbatim}

\tikzstyle{every picture}+=[remember picture]

\everymath{\displaystyle}

\usepackage[linesnumbered,ruled,vlined]{algorithm2e}

\usepackage{lmodern}
\usepackage{anyfontsize}

\usepackage{hyperref}
\hypersetup{colorlinks=true, linkcolor=blue, citecolor=red, urlcolor=blue  }

\begin{document}

\title{Language Model for Large-Text Transmission in Noisy Quantum Communications}


\author{Yuqi~Li}
\thanks{These authors contributed equally}
\affiliation{College of Mathematics Science, Harbin Engineering University, Nantong Street, Harbin, 150001, Heilongjiang, People's Republic of China}

\author{Zhouhang~Shi}
\thanks{These authors contributed equally}
\affiliation{College of Computer Science and Technology, Harbin Engineering University, Nantong Street, Harbin, 150001,
Heilongjiang, People's Republic of China}

\author{Haitao~Ma}
\email{hmamath@hrbeu.edu.cn}
\affiliation{College of Mathematics Science, Harbin Engineering University, Nantong Street, Harbin, 150001, Heilongjiang, People's Republic of China}

\author{Li~Shen}
\email{mathshenli@gmail.com}
\affiliation{School of Cyber Science and Technology, Shenzhen Campus, Sun Yat-sen University, Shenzhen 518107, China}

\author{Jinge~Bao }
\email{jinge.bao@ed.ac.uk}
\affiliation{School of Informatics, University of Edinburgh, Edinburgh EH8 9AB, United Kingdom}
\affiliation{Centre for Quantum Technologies, National University of Singapore, Singapore 138632, Republic of Singapore}
\affiliation{A*STAR Quantum Innovation Centre (Q.InC), Agency for Science, Technology and Research (A*STAR), 2 Fusionopolis Way, Innovis \#08-03, Singapore, 138634, Republic of Singapore}
\affiliation{ Institute of High Performance Computing (IHPC), Agency for Science, Technology and Research (A*STAR), 1 Fusionopolis Way, Connexis \#16-16, Singapore, 138632, Republic of Singapore}

\author{Yunlong~Xiao}
\email{xiao\_yunlong@ihpc.a-star.edu.sg}
\affiliation{A*STAR Quantum Innovation Centre (Q.InC), Agency for Science, Technology and Research (A*STAR), 2 Fusionopolis Way, Innovis \#08-03, Singapore, 138634, Republic of Singapore}
\affiliation{ Institute of High Performance Computing (IHPC), Agency for Science, Technology and Research (A*STAR), 1 Fusionopolis Way, Connexis \#16-16, Singapore, 138632, Republic of Singapore}


\begin{abstract}
Quantum communication has the potential to revolutionize information processing, providing unparalleled security and increased capacity compared to its classical counterpart by using the principles of quantum mechanics. However, the presence of noise remains a major barrier to realizing these advantages. While strategies like quantum error correction and mitigation have been developed to address this challenge, they often come with substantial overhead in physical qubits or sample complexity, limiting their practicality for large-scale information transfer.
Here, we present an alternative approach: applying machine learning frameworks from natural language processing to enhance the performance of noisy quantum communications, focusing on superdense coding. By employing bidirectional encoder representations from transformers (BERT), a model known for its capabilities in natural language processing, we demonstrate improvements in information transfer efficiency without resorting to conventional error correction or mitigation techniques. These results mark a step toward the practical realization of a scalable and resilient quantum internet.
\end{abstract}

\maketitle


\section{Introduction}
In our increasingly interconnected world, effective communication is essential for facilitating the exchange of information, fostering collaboration, and driving problem-solving across various domains. It serves as the foundation of a modern society, enabling the development of technological landscapes. Entering the spotlight, quantum communication~\cite{Bouwmeester1997,Gisin2007,Cozzolino2019,xing2024teleportation,PhysRevLett.126.010503,PhysRevLett.125.230501,xing2023fundamentallimitationscommunicationquantum} is emerging as a revolutionary field that uses the principles of quantum mechanics to achieve unprecedented levels of security and efficiency, signaling a new era in data transmission. Nowadays, quantum information can already be transmitted from satellites to the ground~\cite{ren2017ground,Liao2017,RevModPhys.94.035001,li2024microsatellitebasedrealtimequantumkey}, and quantum networks are being established within cities~\cite{yin2012quantum,Liu2024,Knaut2024}. One notable example of the advantages of quantum communication is  superdense coding~\cite{PhysRevLett.83.3081,PhysRevLett.92.187901,Barreiro2008,Hu2018,li2024communication}, which enables a single qubit channel to transmit the equivalent of two classical bits. This is made possible through shared entanglement between the sender and receiver, effectively doubling the amount of information transmitted. 

The next step in advancing quantum communication involves utilizing the superdense coding protocol as a subroutine for transmitting large volumes of text. However, two practical issues need to be addressed. The first, which is relatively straightforward, involves converting text into a bit string suitable for superdense coding -- this can be easily accomplished using standard ASCII code. The second, more complex challenge arises from the influence of noise on quantum communications, which can degrade performance. Current strategies, including quantum error correction and mitigation, have their pros and cons. For instance, quantum error correction~\cite{PhysRevA.55.900,Aoki2009,RevModPhys.87.307,Krinner2022,Sivak2023,Liu2023,Liu2024a} can efficiently handle errors but often requires additional physical qubits to introduce redundancy, increasing hardware requirements. On the other hand, quantum error mitigation~\cite{PhysRevX.8.031027,RevModPhys.95.045005,PhysRevLett.131.210602,Takagi2022,liu2024nonmarkoviannoisesuppressionsimplified,Liao2025} necessitates multiple experimental repetitions to minimize the impact of errors, potentially undermining the advantages offered by superdense coding.  In light of these challenges, is there an alternative protocol that can effectively mitigate noise in quantum communication without requiring extra physical qubits or repeated communication rounds?

Here, we address this question by introducing a language-model-assisted quantum communication protocol which benefits from both superdense coding and natural language processing (NLP)~\cite{NIPS2017_3f5ee243,devlin-etal-2019-bert,bello2023bert,cao2020bilingual,oswald2022spotspam,hakala2019biomedical,chang2021chinese,10.5555/3454287.3454804,zhuang-etal-2021-robustly,Lan2019ALBERTAL}. In particular, we integrate BERT~\cite{devlin-etal-2019-bert}, a pre-trained language model recognized for its ability to understand the context and predict corrections in tasks such as sentiment analysis~\cite{hoang2019aspect,xu2019bert,8995193, Chen2021,prottasha2022transfer} and spell error correction~\cite{zhang2020spelling,tan2020spelling, nguyen2020neural}, into the superdense coding framework, resulting in what we term post quantum communication BERT (PQC-BERT). PQC-BERT incorporates both a word-level repairing module (WLRM) and a sentence-level repairing module (SLRM), significantly enhancing the fidelity of text transmission in quantum communications. We assess the model's performance by comparing the bit error rate, word error rate, and sentence error rate before and after its implementation through numerical experiments, validating its effectiveness in ensuring reliable text communication. Crucially, PQC-BERT operates across arbitrary noise conditions without requiring prior knowledge of the noise model or its parameters, thereby circumventing the overhead of noise characterization. Furthermore, its ability to accurately detect errors enhances its applicability to broader quantum information processing tasks. This language-model-assisted approach provides a practical solution for improving the reliability of large-text quantum communication, offering the potential to accelerate the development of hybrid classical-quantum networks.


\section{Theoretical Framework}\label{sec:method}
In this work, we integrate superdense coding with machine-learning-based NLP to explore the language-model-assisted quantum communications. This modifies the traditional process of transmitting information -- encoding $\to$ noise $\to$ decoding -- into a more robust pipeline: pre-encoding $\to$ encoding $\to$ noise $\to$ decoding $\to$ post-decoding, as illustrated in Fig.~\hyperref[fig:SDC-BERT]{1(a)}. By exploiting quantum entanglement~\cite{RevModPhys.81.865}, superdense coding efficiently doubles classical information transmission capacity, but its practical application is typically limited to bit string inputs. To overcome this and enable large-scale text transmission, we introduce a pre-encoding step that converts text into 8-bit ASCII code, ensuring compatibility. Following transmission, which may be degraded by noise, a post-decoding phase employs NLP techniques to recover and refine the information. This approach mitigates the impact of noise and enhances the overall reliability and efficiency of superdense-coding-based communications.


\subsection{Quantum Superdense Coding}
To set the stage, let's first examine entanglement -- a core quantum feature that differentiates quantum mechanics from classical physics. This phenomenon, as highlighted in~\cite{RevModPhys.81.865}, fuels revolutionary technologies like quantum computing and communication. In qudit systems, maximally entangled states, commonly known as generalized Bell states, play a fundamental role and are defined as follows
\begin{align}
    \ket{\Phi_{zx}}:= (Z(z)X(x) \otimes I)\ket{\Phi_{00}},
    \quad  
    \forall z,x \in \{0,\ldots, d-1\}.
\end{align}
Here, the state $\Phi_{00}$ is given by 
\begin{align}
    \ket{\Phi_{00}}
    := 
    \frac{1}{\sqrt{d}}\sum_{k=0}^{d-1}{\ket{kk}},
\end{align}
and the Heisenberg-Weyl operators are expressed as~\cite{khatri2024principlesquantumcommunicationtheory}
\begin{align}
    Z(z) &= \sum_{k=0}^{d-1}e^{\frac{2\pi\mathrm{i}kz}{d}}\ketbra{k}{k},\\
    X(x) &= \sum_{k=0}^{d-1}\ketbra{(k+x)\bmod d}{k}.
\end{align}
\begin{figure*}[ht]
\centering
\includegraphics[width=0.9\textwidth]{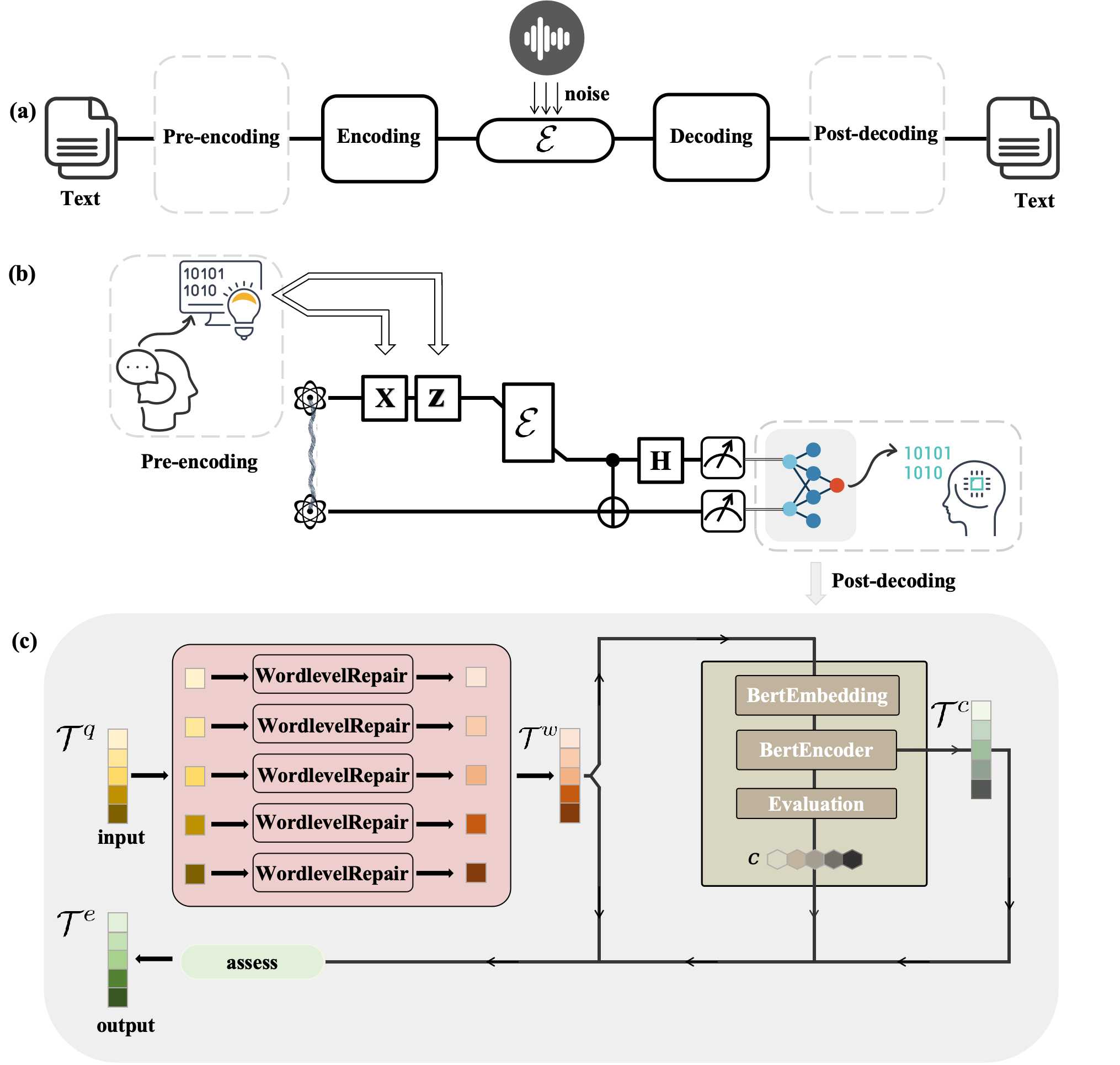}
\caption{{\bf Language-Model-Assisted Quantum Communication.} Figure~(a) shows the outline of our communication protocol designed for large-text transmission tasks. Figure~(b) explicitly illustrates the standard superdense coding. In addition to superdense coding, we encapsulate it with a pre-encoding process and a post-decoding process based on a language model. Figure~(c) further details the structure of our post-decoding component, including the word-level repairing module (WLRM) and the sentence-level repairing module (SLRM).}
\label{fig:SDC-BERT}
\end{figure*}
For qubits, the following four standard Bell states are identified.
\begin{align}
\begin{split}
    \ket{\Phi_{00}} = \frac{1}{\sqrt{2}}(\ket{00}+\ket{11}),\quad
    \ket{\Phi_{01}} = \frac{1}{\sqrt{2}}(\ket{10}+\ket{01}),\\
    \ket{\Phi_{10}} = \frac{1}{\sqrt{2}}(\ket{00}-\ket{11}),\quad
    \ket{\Phi_{11}} = \frac{1}{\sqrt{2}}(\ket{01}-\ket{10}).
\end{split}
\end{align}
These generalized Bell states $\Phi_{zx}$ naturally give rise to projective measurements in the form of $\{\ketbra{\Phi_{zx}}{\Phi_{zx}}\}$, resulting in outcomes $zx$~\cite{PhysRevLett.96.190501}.

With the generalized Bell state at our disposal, we can now introduce quantum superdense coding -- a protocol that enables the transmission of classical information using fewer qubits, provided the sender and receiver share a pre-established entangled resource.  In the qubit case, if the parties pre-share a maximally entangled state $[qq]$, and the sender has access to a noiseless qubit channel $[q\to q]$ to transmit a qubit, it is possible to communicate $2$ classical bits of information. In other words, this simulates the process of $2[c\to c]$
\begin{align}\label{eq:SDC}
    [qq]+[q\to q]\geqslant 2[c\to c].
\end{align}
In practice, communication is often noisy, and the noiseless channel $[q\to q]$ has been replaced by a noisy channel $\mE$ (see Fig.~\hyperref[fig:SDC-BERT]{1(b)}). The formal procedure is outlined as follows:

\begin{enumerate}
    \item \textbf{Entanglement Distribution.} A third party prepared the maximally entangled state $\ket{\Phi_{00}}$ and distributed it to the sender, Alice, and the receiver, Bob. 
    \item \textbf{Encoding.} Alice encodes two bits of classical information $zx\in\{0, 1\}^2$ into one of the Bell states $\ket{\Phi_{zx}}$ by applying Pauli operations $Z(z)X(x)$ to her qubit.
    \item \textbf{Noisy Communication.} Alice then sends her qubit to Bob through the noisy quantum channel $\mE$.
    \item \textbf{Decoding.} Upon receiving the qubit from Alice, Bob performs a Bell measurement. Based on the measurement outcome, Bob decodes the classical information $zx$ that Alice encoded.
\end{enumerate}

A schematic representation of the superdense coding protocol is provided in Fig.~\hyperref[fig:SDC-BERT]{1(b)}. By substituting the Pauli gates with Heisenberg-Weyl operators, we can readily extend the protocol from the qubit case to the general qudit case. For further details and an in-depth discussion on noisy quantum communication, refer to Appendix~\ref{sec:appendix-A}.


\subsection{Pre-Encoding Processing}
To transmit text using a subroutine that incorporates superdense coding, we first convert the characters into a bit string using the standard $8$-bit ASCII code. We refer to this process as ``ASCII encoding'' and denote it by $\mA$. The encoded bit string is then processed through the superdense coding protocol. We denote the original content that sender Alice wishes to transmit to receiver Bob as $\mT= [w_1, \cdots, w_n]$, where $w_i$ ($i\in[n]:=\{1, \ldots, n\}$) represents a word within the text. After transmitting the bit string through the superdense coding protocol, where a noisy channel $\mE$ has occurred, we obtain an output bit string. To recover the transmitted message, we apply the inverse mapping $\mA^{-1}$ to convert the output bit string back into the message $\mT^{q}= [w_1^{q}, \cdots, w_n^{q}]$. Similar to the original content $\mT$, each $w_i^{q}$ signifies a word, with $n$ representing the total number of words. Additionally, each word $w_i^{q}$ can be expressed as $[l^q_{i1}, l^q_{i2}, \cdots, l^q_{im}]$, consisting of 
$m$ letters.


\subsection{Post-Decoding Processing}
Our newly introduced post-decoding process comprises two main components: word-level repairing module (WLRM) and sentence-level repairing module (SLRM). The first component utilizes a predefined dictionary, denoted as $\mathrm{Dict}$, to rectify misspelled words at the word level. The second component employs a language model to tackle more intricate semantic errors at the sentence level.


\subsubsection{Word-Level Repairing Module (SLRM)}

We begin by introducing our word-level repairing module (WLRM). For a given word $w_i^{q}$, derived from superdense coding and subsequently mapped through the inverse mapping $\mA^{-1}$ of the ``ASCII encoding'', we assess its validity against the predefined dictionary $\mathrm{Dict}$. If $w_i^{q}$ is found in $\mathrm{Dict}$, we redefine $w_i^{q}$ as $w_i^{w}$ and proceed to the next word $w_{i+1}^{q}$. Otherwise, the word is identified as misspelled, prompting us to select candidate corrections $w_i^{w}$ from $\mathrm{Dict}$ based on the following measure
\begin{align} 
    \Delta(w, w_i^{q}) 
    := 
    \sum_{j=1}^{m} 
    \mathrm{Hamming}
    \left(\mA(l_{j}), \mA(l_{ij}^{q})\right), 
\end{align} 
where $\mathrm{Hamming}(\cdot,\cdot)$ denotes the Hamming distance, and $l_j$ represents the $j$-th letter of the word $w$. Based on this distance, our objective is to find a word in $\mathrm{Dict}$ that minimizes the aforementioned distance, which we denote as
\begin{align} 
w_i^{w}:= \argmin_{w \in \mathrm{Dict}} \Delta(w, w_i^{q}).
\end{align} 
By repeatedly applying this process, we obtain a word-level repaired text $\mT^{w}= [w_1^{w}, \cdots, w_n^{w}]$. This text 
$\mT^{w}$ is subsequently fed into the next sentence-level repairing module (SLRM). It is worth noting that using spell-corrected text as tokenized input can enhance the correction accuracy of the BERT-based module.


\subsubsection{Sentence-Level Repairing Module (SLRM)}

Our sentence-level repairing module (SLRM) is constructed using BERT and comprises two interconnected networks: the correction network and the evaluation network. The correction network utilizes BERT~\cite{devlin-etal-2019-bert} to analyze the linguistic context effectively. However, we have observed that BERT tends to recommend changes even when the original words are correct. To counter this behavior, we incorporate an evaluation network that  assesses the proposed edits from the correction network, making decisions about whether to accept or reject them.

\begin{figure}[t]
\centering
\includegraphics[width=0.48\textwidth]{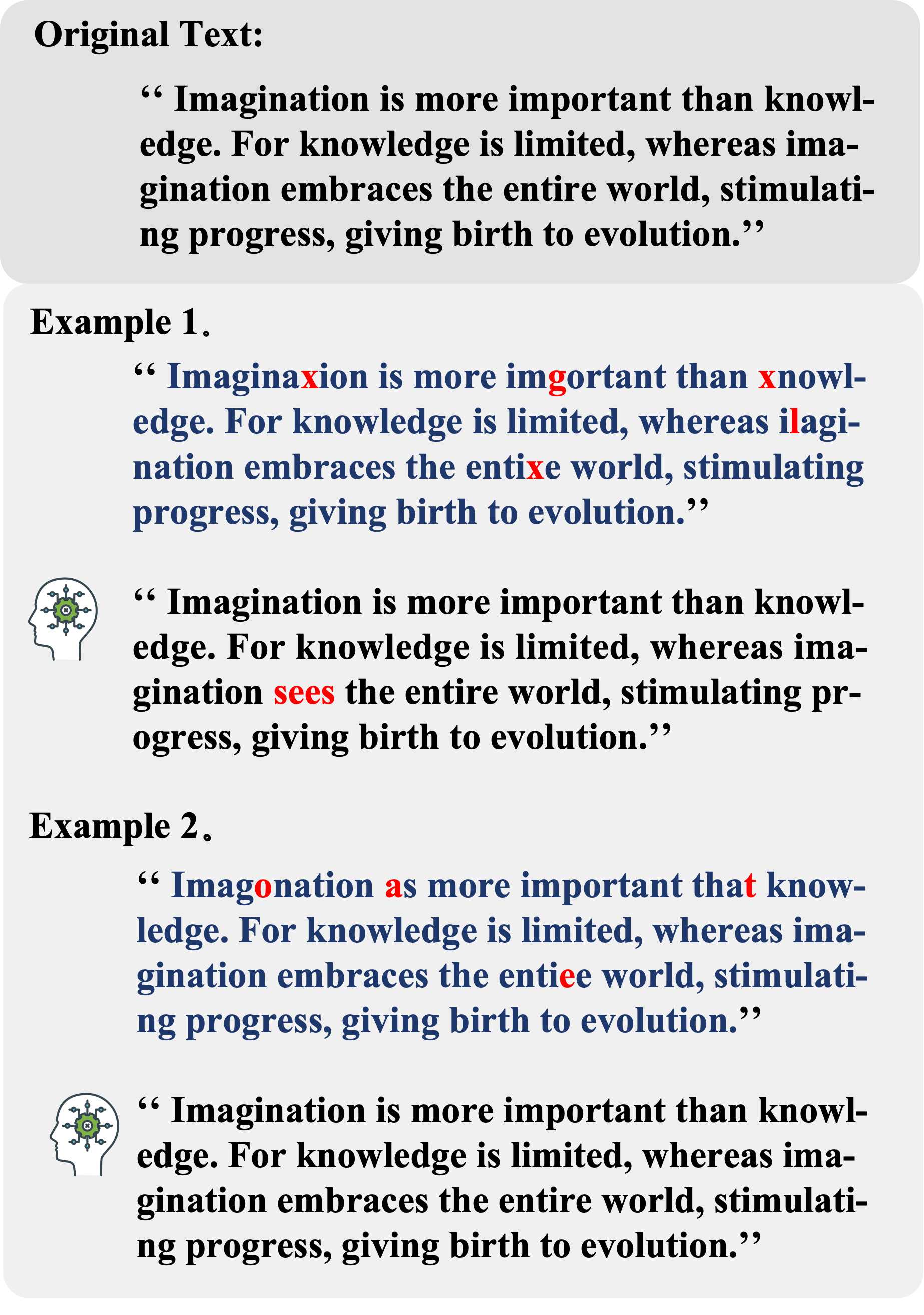}
\caption{{\bf Efficient Text Transmission with PQC-BERT.} 
In each round of noisy communication, errors manifest in different ways. We present two examples, where the text $\mT^{q}$ received after noisy superdense coding is shown in blue, and the corresponding text $\mT^{e}$ processed by our PQC-BERT is shown in black. PQC-BERT successfully corrects nearly all of the errors.
}
\label{fig:text-transmission}
\end{figure}

The correction network employs a sequential multi-class labeling model based on BERT~\cite{devlin-etal-2019-bert}, designed to take corrected text $\mT^{w}$ as input and generate contextually accurate sentences $\mT^{c}$. 
Let the output of BERT be denoted as $P^c = [p^c_1, p^c_2, \cdots, p^c_n]$, where $p^c_i$ is a probability vector. Each element in $p^c_i$ corresponds to the probability of a specific dictionary word appearing in the $i$-th position of the text. For example, if the word `quantum' has the highest probability among all elements in $p^c_1$, the first word of the text will be `quantum.' To construct the full text, we select the word with the highest probability for each position, producing the text $\mT^{c}= [w_1^{c}, \cdots, w_n^{c}]$ (see more details in Appendix~\ref{sec:appendix-B}).

Receiving $\mT^{c}$ as input, the evaluation network produces a vector of confidence scores $c= [c_1, c_2, \ldots, c_n]$. If the score for the $i$-th position, namely $c_i$, is close to $0$, we conclude that the output from BERT is over-corrected; thus, we should select $w_i^{w}$ from the word-level repairing module (WLRM) as the $i$-th word in the text. Conversely, if the score is close to $1$, we trust the correction provided by BERT, indicating that the sentence-level repairing module (SLRM) is necessary, and we take the $i$-th word from the text as $w_i^{c}$. We denote the resulting text from this process as $\mT^{e}$.


\subsubsection{Text Transmission through Noisy Channels}

Here, we employ PQC-BERT to transmit a text and evaluate its performance. Specifically, we consider the ideal text, $\mT$, a quote attributed to Albert Einstein: ``{\it Imagination is more important than knowledge. For knowledge is limited, whereas imagination embraces the entire world, stimulating progress, giving birth to evolution}.'' The text is transmitted through a bit-flip noise channel with a noise probability of 0.01. We compare two versions of the transmitted text: $\mT^{q}$ (without post-decoding correction) and $\mT^{e}$ (with post-decoding correction), as depicted in Fig.~\ref{fig:text-transmission}. The results demonstrate that PQC-BERT successfully corrects nearly all the errors, providing a final text that closely resembles the original.


\section{Performance Analysis}\label{sec:numerical}

In this section, we present numerical experiments to demonstrate the advantages of our language-model-assisted communication protocols, emphasizing the role of machine learning, particularly NLP, in enhancing communication. Our evaluation of the language-model-assisted communication protocols shows that the PQC-BERT module effectively mitigates
noise, significantly reducing the bit error rate, word error
rate, and sentence error rate. These results highlight the superior performance of language-model-assisted quantum communication.


\subsection{Datasets}

Our numerical experiments are based on modified versions of two established datasets: Flickr-8k Text Captioning Dataset (Flickr-8k)~\cite{hodosh2013flickr8k} and Corpus of Linguistic Acceptability (CoLA)~\cite{warstadt2019cola}. The Flickr-8k dataset includes $8,000$ images, each paired with $5$ distinct captions that effectively describe the key entities and events within the images. In contrast, CoLA serves as a benchmark for single-sentence classification, containing sentences sourced from $23$ linguistic publications, expertly annotated for acceptability by their original authors. To assess the performance of our models on text transmission tasks, we manually extracted $12,459$ corrected English sentences from Flickr-8k and $9,078$ from CoLA, resulting in two new datasets: Mini-Flickr and Mini-CoLA.


\subsection{Classical vs Quantum}

As a machine learning framework for natural language processing (NLP), PQC-BERT can effectively mitigate errors in transmitting information through classical channels, making quantum superdense coding seem redundant. But is this truly the case? Beyond increasing the amount of transmitted information, could quantum entanglement also enhance PQC-BERT's ability to mitigate errors? In this study, we show that entanglement does indeed improve PQC-BERT's performance. Even though PQC-BERT doesn't directly interact with quantum resources, we find that its error mitigation is more effective when quantum entanglement is used, under the same bit-flip error and noise strength.

Specifically, given a noisy channel $\mE$ and an ideal text $\mT$, we simulate the action of $\mE$ to generate a corrupted version, then apply WLRM to produce a corrected text $\mT^w$ through word-level correction. During training, $\mT^w$ serves as the input and $\mT$ as the target output for the neural network. In testing, we assess the similarity between the original ideal text $\mT$ (before noise) and the final output after the post-decoding process. This evaluation captures the overall performance of the language-model-assisted communication protocol.

\begin{table}[t]
\begin{center}
\resizebox{0.5\textwidth}{!}{
\begin{tabular}{cccccc}
\toprule
\multicolumn{1}{c}{\bf Dataset}  &\multicolumn{1}{c}{\bf Channel} &\multicolumn{1}{c}{\bf Accuracy} &\multicolumn{1}{c}{\bf Precision} &\multicolumn{1}{c}{\bf Recall} &\multicolumn{1}{c}{\bf F1-Score}\\ 
\hline \hline \addlinespace
\multirow{2}{*}{Mini-Flickr} & Classical & 0.6336 & 0.6583 & 0.6167 & 0.7045 \\
& Quantum & \textbf{0.7642} & \textbf{0.7971}& \textbf{0.7491} & \textbf{0.7725} \\
\addlinespace \hline \addlinespace
\multirow{2}{*}{Mini-CoLA} & Classical & 0.5104 & 0.7108 & 0.5903 & 0.5172 \\
& Quantum & \textbf{0.7401} & \textbf{0.7971} & \textbf{0.6611} & \textbf{0.6765} \\ 
\addlinespace \bottomrule
\end{tabular}
}
\end{center}
\caption{{\bf PQC-BERT-Assisted Classical vs Quantum Communication.} 
PQC-BERT was applied to classical and quantum (qubit-based) communication protocols, both subject to a fixed bit-flip noise rate of $0.01$. The results demonstrate that, despite identical noise conditions, PQC-BERT-assisted quantum communication achieves superior performance compared to the classical case, illustrating the potential benefits of leveraging quantum resources.
}
\label{table:Performance}
\end{table}

To ensure reliability, we conduct independent and replicated experiments, reporting the average performance across four standard metrics: {\it Accuracy}, {\it Recall}, {\it Precision}, and {\it F1-score}. Detailed definitions of these metrics are provided in Appendix~\ref{sec:appendix-E}. As summarized in Table~\ref{table:Performance}, we present the performance of PQC-BERT under a bit-flip noise rate of $0.01$ for both quantum and classical communications. The results demonstrate that PQC-BERT remains highly effective and robust in the presence of noise. Furthermore, we observe that PQC-BERT’s performance improves when assisted by entanglement, specifically via the superdense coding protocol, highlighting the broader potential of quantum resources in enhancing machine learning capabilities.


\subsection{Qubit vs Qudit}

\begin{figure*}[hptb]
\centering
\includegraphics[width=1\textwidth]{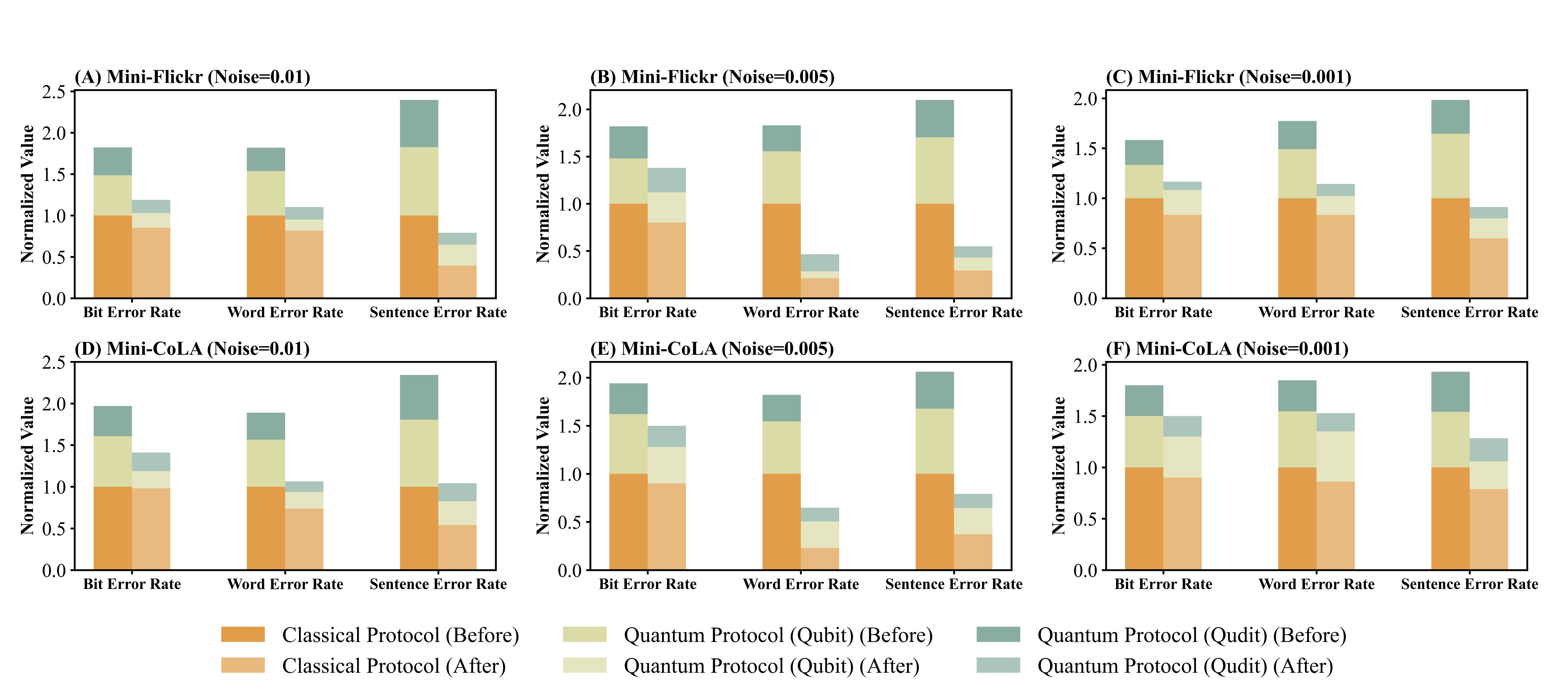}
\caption{{\bf PQC-BERT-Assisted Qubit vs Qudit Superdense Coding.} 
The performance of qubit- and qudit-based ($d=4$) superdense coding is evaluated under various noise parameters of the bit-flip channel, with classical communication included as a benchmark across different datasets.
}
\label{fig:classical-quantum}
\end{figure*}

We proceed by evaluating the performance of the language-model-assisted communication protocol under bit-flip noise, comparing the results when a maximally entangled qubit state is replaced by a maximally entangled qudit state (with $d=4$). This allows us to assess whether the qudit protocol yields better performance. For comparison, we also include the case of direct transmission through a classical channel as a benchmark. The results are shown in Fig.~\ref{fig:classical-quantum}, and our analysis focuses on three key metrics:

\begin{enumerate}
    \item \textbf{Bit Error Rate (BER).} This metric quantifies the ratio of erroneous bits to the total number of bits transmitted. It is affected by the channel's noise parameters, providing valuable insights into the system's sensitivity to noise, its interference suppression capabilities, and its overall communication capacity -- making it a crucial metric in information theory.
    \item \textbf{Word Error Rate (WER).} This metric assesses the proportion of errors within transmitted data blocks, defined as the number of erroneous words divided by the total number of transmitted words. It effectively evaluates the quality of transmission at the word level.
    \item \textbf{Sentence Error Rate (SER).} This metric measures the performance of our text communication protocol by calculating the ratio of erroneous sentences to the total number of sentences transmitted. It provides valuable insight into the protocol's accuracy at the sentence level.
\end{enumerate}

Figure~\ref{fig:classical-quantum} visualizes the numerical experimental data, comparing the performance of classical communication, qubit-based superdense coding, and qudit-based superdense coding, both before and after applying PQC-BERT. In the figure, error rates are proportionally rescaled so that the maximum value for each metric is normalized to 1, with taller bars indicating poorer performance (higher error rates) and shorter bars indicating better performance (lower error rates). Across both datasets and for each metric under fixed noise parameters, quantum communication protocols -- whether using qubits or qudits ($d=4$) -- consistently outperform classical communication, with qudit-based superdense coding further surpassing the qubit-based case. This observation suggests that increased entanglement leads to improved transmission performance. Overall, the results highlight that quantum superdense coding not only achieves higher communication capacity but also exhibits improved resilience to noise when assisted by PQC-BERT.

\begin{figure*}
\centering
\includegraphics[width=1\textwidth]{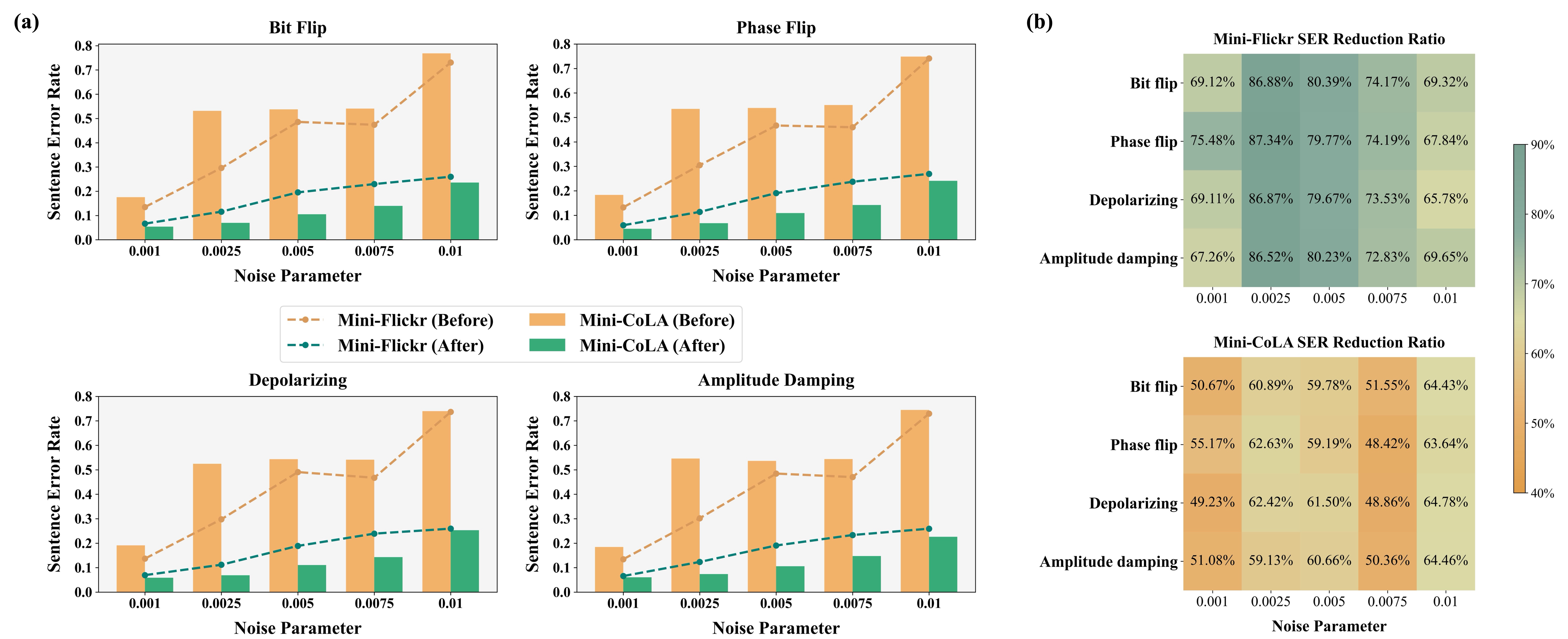}
\caption{{\bf Performance Profiling Across Varied Noise Models.} 
Figure (a) shows the sentence error rate (SER) before and after PQC-BERT, indicated by orange and green bars/lines, respectively. Solid bars correspond to results on Mini-Flickr, while dashed lines represent Mini-CoLA. Figure (b) reports the SER reduction ratio, as defined in Eq.~\ref{eq:SER}.
}
\label{fig:different-noise}
\end{figure*}

In large-scale text transmission, the sentence error rate is a critical metric, as it directly impacts the readability of the received message after noisy communication. While detailed numerical results are presented in Appendix~\ref{sec:appendix-F}, we briefly illustrate the effectiveness of our language model-assisted quantum communication protocol here. Using a bit-flip channel with noise strength $0.01$ to transmit qubit states, we observe notably high sentence error rates: $76.85\%$ for Mini-Flickr and $73.06\%$ for Mini-CoLA. Despite the relatively low noise level, these error rates severely degrade readability, rendering most sentences unintelligible. After incorporating the PQC-BERT module, the sentence error rates are dramatically reduced to $23.58\%$ for Mini-Flickr and $25.99\%$ for Mini-CoLA, enabling the receiver, Bob, to successfully reconstruct and comprehend the transmitted text.


\subsection{Performance Across Different Noise Models}

In classical communication, bit-flip errors represent the primary source of transmission degradation. In quantum communication, however, a broader and more complex range of error mechanisms must be addressed. To evaluate the versatility of PQC-BERT, we tested its performance across four distinct quantum noise models: bit-flip, phase-flip, depolarizing, and amplitude-damping noise. Sentences encoded in qubit-based quantum systems were transmitted through each noise channel, using the Mini-Flickr and Mini-CoLA datasets to generate distinct error patterns for training. Notably, PQC-BERT was trained without explicit access to noise model labels or parametric information. This setting allows a stringent assessment of whether the model can autonomously correct diverse quantum errors without prior knowledge of the underlying noise processes.

The results are summarized in Fig.~\ref{fig:different-noise}. Fig.~\hyperref[fig:different-noise]{4(a)} illustrates the reduction in SER before and after correction, with four different noise exhibiting consistent trends under identical noise parameters. For a noise parameter of 0.01, the SER for the four noise models before correction ranges from $75.41\% \pm 1.44\%$ for Mini-Flickr and $73.52\% \pm 0.57\%$ for Mini-CoLA. After correction, the SER decreases to $23.95\% \pm 1.35\%$ and $26.43\% \pm 0.50\%$, respectively, indicating the robustness of the model across various noise models.
Fig.~\hyperref[fig:different-noise]{4(b)} extends this analysis by presenting the error reduction ratio, defined as:
\begin{align}\label{eq:SER}
\text{ SER Reduction Ratio} = \frac{\text{SER}_{\text{Before}} - \text{SER}_{\text{After}}}{\text{SER}_{\text{Before}}}.   
\end{align}
This metric quantifies the relative decrease in the SER, where a higher value indicates a more effective error mitigation.
The results clearly underscore the pivotal role of PQC-BERT in enhancing communication performance and demonstrate its strong adaptability to varying noise conditions.

\begin{figure}
\centering
\includegraphics[width=0.48\textwidth]{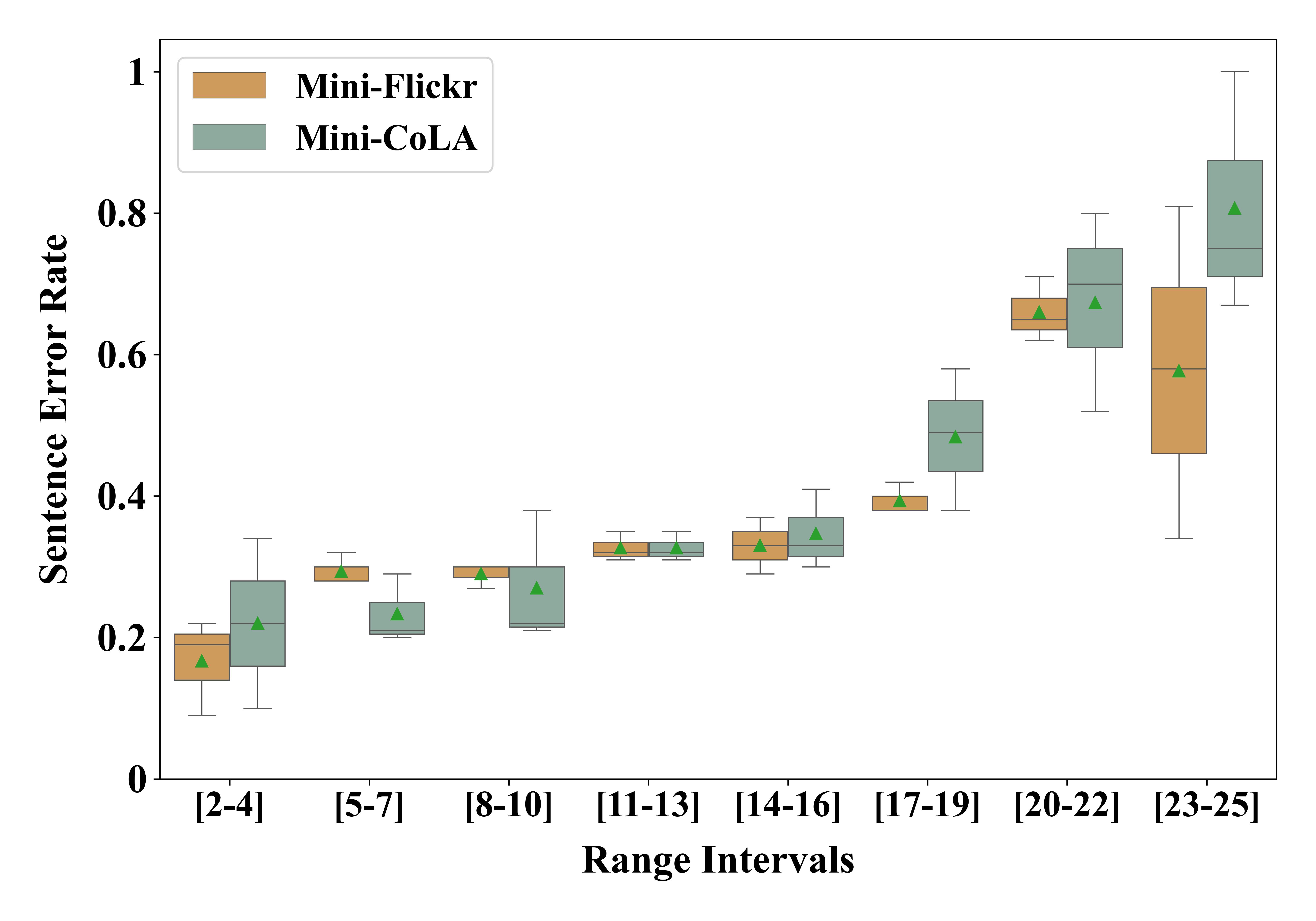}
\caption{{\bf Impact of Sentence Length on Performance.} Dependence of sentence error rate (SER) on sentence length across two datasets, evaluated under a bit-flip noise model with noise rate $0.01$.
}
\label{fig:length}
\end{figure}


\subsection{Sentence Length and Its Effect}

PQC-BERT maintains consistent performance across diverse noise models, highlighting its potential for enhancing quantum communication. Nonetheless, variability between the Mini-Flickr and Mini-CoLA datasets suggests a sensitivity to sentence length. To probe this effect, we analyzed performance as a function of sentence length and found that PQC-BERT operates optimally on sentences containing 5–16 words (corresponding to 100–320 qubits; Fig.~\ref{fig:length}). Guided by this observation, we partition information into transmission units of up to 16 words, aligning with the model's peak correction capability. This unit-based strategy not only maximizes performance but also establishes a standardized framework for future deployments.
Representative examples of unit-based and overall transmission performance are provided in Appendix~\ref{sec:appendix-G}.

PQC-BERT's optimal performance on sentences containing 5–16 words reflects the quadratic scaling of the self-attention mechanism with sequence length. Shorter sequences enable more efficient computation, allowing the model to capture token relationships more effectively. Extending PQC-BERT's capabilities to longer sequences will require the integration of global feature modeling and the adoption of architectures optimized for extended contexts. In particular, Longformer~\cite{beltagy2020longformer}, which employs sparse attention to efficiently capture long-range dependencies, and BigBird~\cite{zaheer2021bigbirdtransformerslonger}, which combines global, local, and random attention patterns, offer promising pathways. Incorporating such architectures could mitigate the performance degradation observed for longer inputs and enhance the model's reliability in quantum communication tasks involving complex or extensive information transmission. A detailed investigation along these lines is beyond the scope of the present work and is left for future exploration.

\begin{figure*}
\centering
\includegraphics[width=1\textwidth]{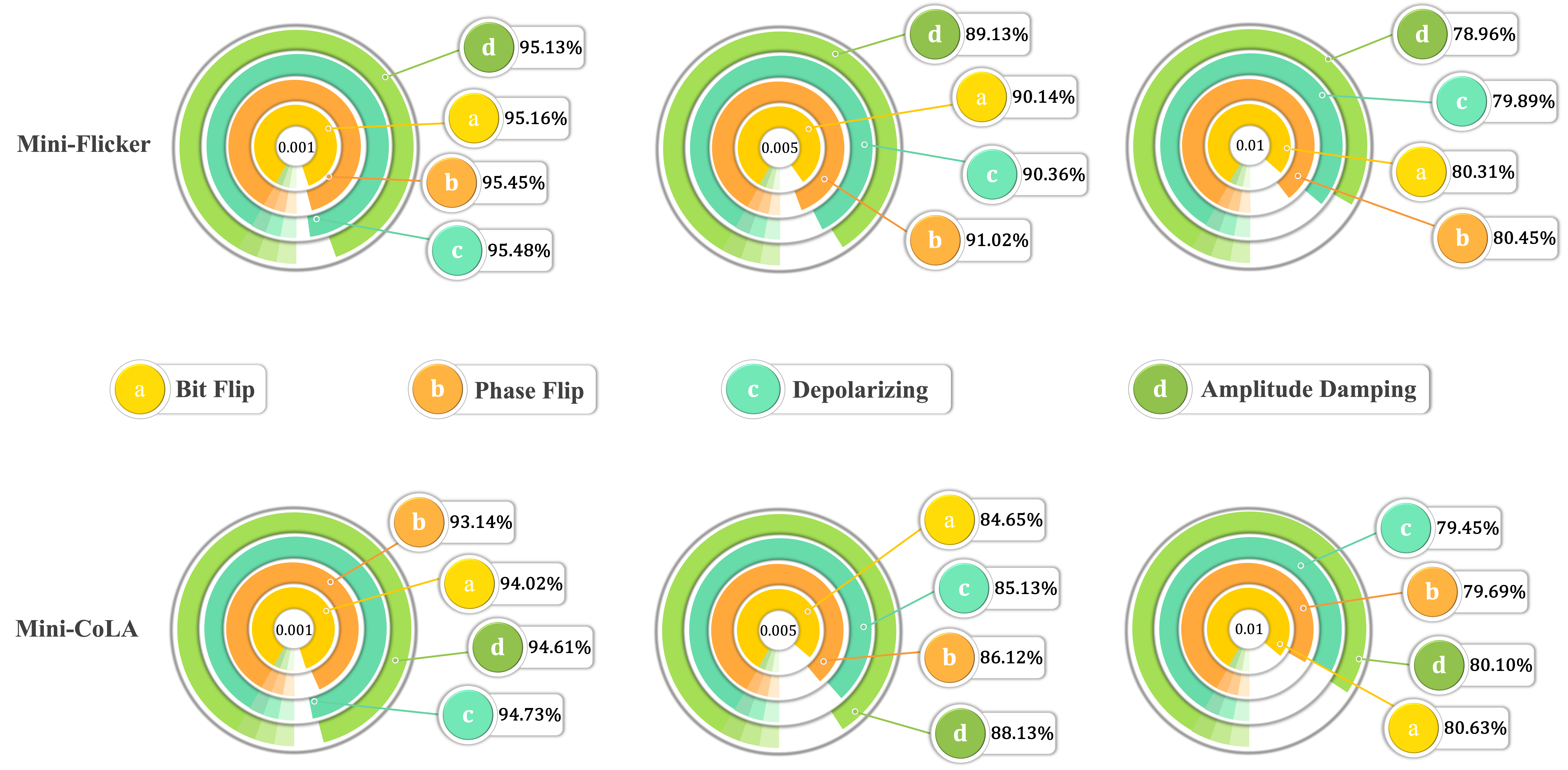}
\captionsetup{justification=raggedright, singlelinecheck=false}
\caption{{\bf Error Detection Under Varying Noise Models.} 
Each circle displays a noise parameter at its center, with colors indicating the different noise types. Figure (a) depicts the probability of successful error detection for Mini-Flickr, while Figure (b) presents the corresponding result for Mini-CoLA.
}
\label{fig:error-detection}
\end{figure*}


\subsection{Error Detection}\label{sec:application}

Detecting and localizing errors are foundational tasks in both classical and quantum information processing, typically forming the first step toward error correction~\cite{10313604}. In quantum error correction, detection proceeds through the syndrome bits, which identify the location of errors and enable their systematic correction. However, accurate localization demands additional ancillary systems, introducing a nontrivial overhead of quantum resources. In large-text transmission, where noise inevitably degrades communication, a natural question arises: can our framework similarly detect and localize errors efficiently? We answer affirmatively. As shown in Fig.~\ref{fig:error-detection}, PQC-BERT exhibits high accuracy in pinpointing error locations across a variety of noise models.

Nonetheless, perfect reliability remains out of reach, as PQC-BERT occasionally prioritizes the extraction of global contextual features over the precise localization of individual errors, leading to a finite failure probability. Nevertheless, its high accuracy and efficiency in error detection and localization across a broad range of noise models remain compelling. These strengths suggest that PQC-BERT could be integrated with conventional quantum error correction protocols, serving as a context-sensitive processor to rapidly flag likely errors. Such hybrid strategies, which combine machine-learned error localization with structured correction mechanisms, may offer a more favorable balance between resource efficiency, accuracy, and fault tolerance -- an avenue that merits further exploration.


\section{Discussions}

Quantum communications offer significant advantages over classical protocols, particularly in capacity and privacy. However, these benefits are often undermined by noise, which is nearly unavoidable in practical applications. In this work, we leverage the power of language models in machine learning to bolster the performance of a key quantum communication protocol -- quantum superdense coding -- in the presence of noise. We present PQC-BERT, a model that extends conventional superdense coding to enable the transmission of text while effectively mitigating errors that arise during quantum communication. Distinct from conventional quantum error correction techniques, our method protects information without the need for additional systems. Furthermore, unlike typical quantum error mitigation strategies, PQC-BERT operates without requiring extra samples to address noise effects. This makes our approach both resource-efficient and easy to implement, highlighting the remarkable potential of classical machine learning techniques in advancing quantum communications.

The fusion of classical machine learning with quantum superdense coding also opens up new avenues for exploration and raises intriguing questions for future exploration.
First, superdense coding is typically used to transmit classical bit strings through quantum communication, demonstrating its advantage in conveying more information per channel use. A key challenge now lies in extending the protocol to support more complex data types, such as images and audio. Given the distinct structure and features of such data, more advanced pre-encoding schemes -- beyond the ``ASCII encoding'' -- may be needed to fully optimize the transmission process. Developing these schemes could unlock even broader applications of quantum communication protocols in the future. 
Second, while our focus has been on language models and superdense coding, quantum communication encompasses a wider array of tasks. As we move towards building a quantum internet~\cite{carrasquilla2020machine,melko2019restricted,carleo2017solving,carrasquilla2017machine}, there is exciting potential to integrate artificial intelligence with other key protocols~\cite{PRXQuantum.1.010301,biamonte2017quantum,dunjko2018machine,guo2024quantum,yu2023provable,Cherrat_2024,tian2023recent,wang2023symmetric}, such as quantum teleportation and quantum repeaters, to further push the boundaries of this field. 
Third, in the field of natural language processing, numerous models have emerged following the BERT architecture. Will these subsequent models outperform BERT in the post-decoding phase of superdense coding? Could different language models reveal distinct advantages for different noise models? Our numerical experiments are based on processed datasets, Mini-Flickr and Mini-CoLA, masked with simulated quantum noise. It will be beneficial to further evaluate the performance of language-model-assisted quantum communications in physical systems, such as photonics~\cite{10.1063/1.5115814,RevModPhys.92.035005,Yuan2023}. However, exploring this aspect falls beyond the scope of the current work and will be addressed in future research.

\begin{acknowledgments}
We would like to thank Junjing Xing, Chenghong Zhu, Jun Yong Khoo, Jian Feng Kong, Jun Ye, and Lorcan Conlon for fruitful discussions. 
This research is supported by A*STAR under C23091703 and Q.InC Strategic Research and Translational Thrust.
Yunlong Xiao acknowledges support from A*STAR under its Central Research Funds and Career Development Fund (C243512002). 
Jinge Bao is supported by Quantum Advantage Pathfinder (QAP) project of UKRI Engineering and Physical Sciences Research Council (EP/X026167/1).
Yuqi Li, Zhouhang Shi, and Haitao Ma are Supported by the Stable Supporting Fund of National Key Laboratory of Underwater Acoustic Technology (KY12400230004). 
\end{acknowledgments}


\bibliography{References.bib}

\begin{thebibliography}{85}%
\makeatletter
\providecommand \@ifxundefined [1]{%
 \@ifx{#1\undefined}
}%
\providecommand \@ifnum [1]{%
 \ifnum #1\expandafter \@firstoftwo
 \else \expandafter \@secondoftwo
 \fi
}%
\providecommand \@ifx [1]{%
 \ifx #1\expandafter \@firstoftwo
 \else \expandafter \@secondoftwo
 \fi
}%
\providecommand \natexlab [1]{#1}%
\providecommand \enquote  [1]{``#1''}%
\providecommand \bibnamefont  [1]{#1}%
\providecommand \bibfnamefont [1]{#1}%
\providecommand \citenamefont [1]{#1}%
\providecommand \href@noop [0]{\@secondoftwo}%
\providecommand \href [0]{\begingroup \@sanitize@url \@href}%
\providecommand \@href[1]{\@@startlink{#1}\@@href}%
\providecommand \@@href[1]{\endgroup#1\@@endlink}%
\providecommand \@sanitize@url [0]{\catcode `\\12\catcode `\$12\catcode
  `\&12\catcode `\#12\catcode `\^12\catcode `\_12\catcode `\%12\relax}%
\providecommand \@@startlink[1]{}%
\providecommand \@@endlink[0]{}%
\providecommand \url  [0]{\begingroup\@sanitize@url \@url }%
\providecommand \@url [1]{\endgroup\@href {#1}{\urlprefix }}%
\providecommand \urlprefix  [0]{URL }%
\providecommand \Eprint [0]{\href }%
\providecommand \doibase [0]{https://doi.org/}%
\providecommand \selectlanguage [0]{\@gobble}%
\providecommand \bibinfo  [0]{\@secondoftwo}%
\providecommand \bibfield  [0]{\@secondoftwo}%
\providecommand \translation [1]{[#1]}%
\providecommand \BibitemOpen [0]{}%
\providecommand \bibitemStop [0]{}%
\providecommand \bibitemNoStop [0]{.\EOS\space}%
\providecommand \EOS [0]{\spacefactor3000\relax}%
\providecommand \BibitemShut  [1]{\csname bibitem#1\endcsname}%
\let\auto@bib@innerbib\@empty
\bibitem [{\citenamefont {Bouwmeester}\ \emph {et~al.}(1997)\citenamefont
  {Bouwmeester}, \citenamefont {Pan}, \citenamefont {Mattle},\ and\
  \citenamefont {et~al.}}]{Bouwmeester1997}%
  \BibitemOpen
  \bibfield  {author} {\bibinfo {author} {\bibfnamefont {D.}~\bibnamefont
  {Bouwmeester}}, \bibinfo {author} {\bibfnamefont {J.~W.}\ \bibnamefont
  {Pan}}, \bibinfo {author} {\bibfnamefont {K.}~\bibnamefont {Mattle}},\ and\
  \bibinfo {author} {\bibnamefont {et~al.}},\ }\bibfield  {title} {\bibinfo
  {title} {Experimental quantum teleportation},\ }\href
  {https://doi.org/https://doi.org/10.1038/37539} {\bibfield  {journal}
  {\bibinfo  {journal} {Nature}\ }\textbf {\bibinfo {volume} {390}},\ \bibinfo
  {pages} {575} (\bibinfo {year} {1997})}\BibitemShut {NoStop}%
\bibitem [{\citenamefont {Gisin}\ and\ \citenamefont {Thew}(2007)}]{Gisin2007}%
  \BibitemOpen
  \bibfield  {author} {\bibinfo {author} {\bibfnamefont {N.}~\bibnamefont
  {Gisin}}\ and\ \bibinfo {author} {\bibfnamefont {R.}~\bibnamefont {Thew}},\
  }\bibfield  {title} {\bibinfo {title} {Quantum communication},\ }\href
  {https://doi.org/10.1038/nphoton.2007.22} {\bibfield  {journal} {\bibinfo
  {journal} {Nature Photonics}\ }\textbf {\bibinfo {volume} {1}},\ \bibinfo
  {pages} {165} (\bibinfo {year} {2007})}\BibitemShut {NoStop}%
\bibitem [{\citenamefont {Cozzolino}\ \emph {et~al.}(2019)\citenamefont
  {Cozzolino}, \citenamefont {Da~Lio}, \citenamefont {Bacco},\ and\
  \citenamefont {Oxenløwe}}]{Cozzolino2019}%
  \BibitemOpen
  \bibfield  {author} {\bibinfo {author} {\bibfnamefont {D.}~\bibnamefont
  {Cozzolino}}, \bibinfo {author} {\bibfnamefont {B.}~\bibnamefont {Da~Lio}},
  \bibinfo {author} {\bibfnamefont {D.}~\bibnamefont {Bacco}},\ and\ \bibinfo
  {author} {\bibfnamefont {L.~K.}\ \bibnamefont {Oxenløwe}},\ }\bibfield
  {title} {\bibinfo {title} {High-dimensional quantum communication: Benefits,
  progress, and future challenges},\ }\href
  {https://doi.org/https://doi.org/10.1002/qute.201900038} {\bibfield
  {journal} {\bibinfo  {journal} {Quantum Science and Technology}\ }\textbf
  {\bibinfo {volume} {5}},\ \bibinfo {pages} {043001} (\bibinfo {year}
  {2019})}\BibitemShut {NoStop}%
\bibitem [{\citenamefont {Xing}\ \emph {et~al.}(2024)\citenamefont {Xing},
  \citenamefont {Li}, \citenamefont {Qu}, \citenamefont {Xiao}, \citenamefont
  {Fan}, \citenamefont {Ma}, \citenamefont {Xue}, \citenamefont {Bharti},
  \citenamefont {Koh},\ and\ \citenamefont {Xiao}}]{xing2024teleportation}%
  \BibitemOpen
  \bibfield  {author} {\bibinfo {author} {\bibfnamefont {J.}~\bibnamefont
  {Xing}}, \bibinfo {author} {\bibfnamefont {Y.}~\bibnamefont {Li}}, \bibinfo
  {author} {\bibfnamefont {D.}~\bibnamefont {Qu}}, \bibinfo {author}
  {\bibfnamefont {L.}~\bibnamefont {Xiao}}, \bibinfo {author} {\bibfnamefont
  {Z.}~\bibnamefont {Fan}}, \bibinfo {author} {\bibfnamefont {H.}~\bibnamefont
  {Ma}}, \bibinfo {author} {\bibfnamefont {P.}~\bibnamefont {Xue}}, \bibinfo
  {author} {\bibfnamefont {K.}~\bibnamefont {Bharti}}, \bibinfo {author}
  {\bibfnamefont {D.~E.}\ \bibnamefont {Koh}},\ and\ \bibinfo {author}
  {\bibfnamefont {Y.}~\bibnamefont {Xiao}},\ }\bibfield  {title} {\bibinfo
  {title} {Teleportation with embezzling catalysts},\ }\href
  {https://doi.org/10.1038/s42005-024-01828-x} {\bibfield  {journal} {\bibinfo
  {journal} {Communications Physics}\ }\textbf {\bibinfo {volume} {7}},\
  \bibinfo {pages} {357} (\bibinfo {year} {2024})}\BibitemShut {NoStop}%
\bibitem [{\citenamefont {Hu}\ \emph {et~al.}(2021)\citenamefont {Hu},
  \citenamefont {Huang}, \citenamefont {Sheng}, \citenamefont {Zhou},
  \citenamefont {Liu}, \citenamefont {Guo}, \citenamefont {Zhang},
  \citenamefont {Xing}, \citenamefont {Huang}, \citenamefont {Li},\ and\
  \citenamefont {Guo}}]{PhysRevLett.126.010503}%
  \BibitemOpen
  \bibfield  {author} {\bibinfo {author} {\bibfnamefont {X.-M.}\ \bibnamefont
  {Hu}}, \bibinfo {author} {\bibfnamefont {C.-X.}\ \bibnamefont {Huang}},
  \bibinfo {author} {\bibfnamefont {Y.-B.}\ \bibnamefont {Sheng}}, \bibinfo
  {author} {\bibfnamefont {L.}~\bibnamefont {Zhou}}, \bibinfo {author}
  {\bibfnamefont {B.-H.}\ \bibnamefont {Liu}}, \bibinfo {author} {\bibfnamefont
  {Y.}~\bibnamefont {Guo}}, \bibinfo {author} {\bibfnamefont {C.}~\bibnamefont
  {Zhang}}, \bibinfo {author} {\bibfnamefont {W.-B.}\ \bibnamefont {Xing}},
  \bibinfo {author} {\bibfnamefont {Y.-F.}\ \bibnamefont {Huang}}, \bibinfo
  {author} {\bibfnamefont {C.-F.}\ \bibnamefont {Li}},\ and\ \bibinfo {author}
  {\bibfnamefont {G.-C.}\ \bibnamefont {Guo}},\ }\bibfield  {title} {\bibinfo
  {title} {Long-distance entanglement purification for quantum communication},\
  }\href {https://doi.org/10.1103/PhysRevLett.126.010503} {\bibfield  {journal}
  {\bibinfo  {journal} {Phys. Rev. Lett.}\ }\textbf {\bibinfo {volume} {126}},\
  \bibinfo {pages} {010503} (\bibinfo {year} {2021})}\BibitemShut {NoStop}%
\bibitem [{\citenamefont {Hu}\ \emph {et~al.}(2020)\citenamefont {Hu},
  \citenamefont {Zhang}, \citenamefont {Liu} \emph
  {et~al.}}]{PhysRevLett.125.230501}%
  \BibitemOpen
  \bibfield  {author} {\bibinfo {author} {\bibfnamefont {X.-M.}\ \bibnamefont
  {Hu}}, \bibinfo {author} {\bibfnamefont {C.}~\bibnamefont {Zhang}}, \bibinfo
  {author} {\bibfnamefont {B.-H.}\ \bibnamefont {Liu}}, \emph {et~al.},\
  }\bibfield  {title} {\bibinfo {title} {Experimental high-dimensional quantum
  teleportation},\ }\href {https://doi.org/10.1103/PhysRevLett.125.230501}
  {\bibfield  {journal} {\bibinfo  {journal} {Phys. Rev. Lett.}\ }\textbf
  {\bibinfo {volume} {125}},\ \bibinfo {pages} {230501} (\bibinfo {year}
  {2020})}\BibitemShut {NoStop}%
\bibitem [{\citenamefont {Xing}\ \emph {et~al.}(2023)\citenamefont {Xing},
  \citenamefont {Feng}, \citenamefont {Fan}, \citenamefont {Ma}, \citenamefont
  {Bharti}, \citenamefont {Koh},\ and\ \citenamefont
  {Xiao}}]{xing2023fundamentallimitationscommunicationquantum}%
  \BibitemOpen
  \bibfield  {author} {\bibinfo {author} {\bibfnamefont {J.}~\bibnamefont
  {Xing}}, \bibinfo {author} {\bibfnamefont {T.}~\bibnamefont {Feng}}, \bibinfo
  {author} {\bibfnamefont {Z.}~\bibnamefont {Fan}}, \bibinfo {author}
  {\bibfnamefont {H.}~\bibnamefont {Ma}}, \bibinfo {author} {\bibfnamefont
  {K.}~\bibnamefont {Bharti}}, \bibinfo {author} {\bibfnamefont {D.~E.}\
  \bibnamefont {Koh}},\ and\ \bibinfo {author} {\bibfnamefont {Y.}~\bibnamefont
  {Xiao}},\ }\href@noop {} {\bibinfo {title} {Fundamental limitations on
  communication over a quantum network}} (\bibinfo {year} {2023}),\ \Eprint
  {https://arxiv.org/abs/2306.04983} {arXiv:2306.04983 [quant-ph]} \BibitemShut
  {NoStop}%
\bibitem [{\citenamefont {Ren}\ \emph {et~al.}(2017)\citenamefont {Ren},
  \citenamefont {Xu}, \citenamefont {Yong}, \citenamefont {Zhang},
  \citenamefont {Liao}, \citenamefont {Yin}, \citenamefont {Liu}, \citenamefont
  {Cai}, \citenamefont {Yang}, \citenamefont {Li} \emph
  {et~al.}}]{ren2017ground}%
  \BibitemOpen
  \bibfield  {author} {\bibinfo {author} {\bibfnamefont {J.-G.}\ \bibnamefont
  {Ren}}, \bibinfo {author} {\bibfnamefont {P.}~\bibnamefont {Xu}}, \bibinfo
  {author} {\bibfnamefont {H.-L.}\ \bibnamefont {Yong}}, \bibinfo {author}
  {\bibfnamefont {L.}~\bibnamefont {Zhang}}, \bibinfo {author} {\bibfnamefont
  {S.-K.}\ \bibnamefont {Liao}}, \bibinfo {author} {\bibfnamefont
  {J.}~\bibnamefont {Yin}}, \bibinfo {author} {\bibfnamefont {W.-Y.}\
  \bibnamefont {Liu}}, \bibinfo {author} {\bibfnamefont {W.-Q.}\ \bibnamefont
  {Cai}}, \bibinfo {author} {\bibfnamefont {M.}~\bibnamefont {Yang}}, \bibinfo
  {author} {\bibfnamefont {L.}~\bibnamefont {Li}}, \emph {et~al.},\ }\bibfield
  {title} {\bibinfo {title} {Ground-to-satellite quantum teleportation},\
  }\href {https://doi.org/https://doi.org/10.1038/nature23675} {\bibfield
  {journal} {\bibinfo  {journal} {Nature}\ }\textbf {\bibinfo {volume} {549}},\
  \bibinfo {pages} {70} (\bibinfo {year} {2017})}\BibitemShut {NoStop}%
\bibitem [{\citenamefont {Liao}\ \emph {et~al.}(2017)\citenamefont {Liao},
  \citenamefont {Cai}, \citenamefont {Liu}, \citenamefont {Zhang},
  \citenamefont {Li}, \citenamefont {Ren} \emph {et~al.}}]{Liao2017}%
  \BibitemOpen
  \bibfield  {author} {\bibinfo {author} {\bibfnamefont {S.-K.}\ \bibnamefont
  {Liao}}, \bibinfo {author} {\bibfnamefont {W.-Q.}\ \bibnamefont {Cai}},
  \bibinfo {author} {\bibfnamefont {W.-Y.}\ \bibnamefont {Liu}}, \bibinfo
  {author} {\bibfnamefont {L.}~\bibnamefont {Zhang}}, \bibinfo {author}
  {\bibfnamefont {Y.}~\bibnamefont {Li}}, \bibinfo {author} {\bibfnamefont
  {J.-G.}\ \bibnamefont {Ren}}, \emph {et~al.},\ }\bibfield  {title} {\bibinfo
  {title} {Satellite-to-ground quantum key distribution},\ }\href
  {https://doi.org/10.1038/nature23655} {\bibfield  {journal} {\bibinfo
  {journal} {Nature}\ }\textbf {\bibinfo {volume} {549}},\ \bibinfo {pages}
  {43} (\bibinfo {year} {2017})}\BibitemShut {NoStop}%
\bibitem [{\citenamefont {Lu}\ \emph {et~al.}(2022)\citenamefont {Lu},
  \citenamefont {Cao}, \citenamefont {Peng},\ and\ \citenamefont
  {Pan}}]{RevModPhys.94.035001}%
  \BibitemOpen
  \bibfield  {author} {\bibinfo {author} {\bibfnamefont {C.-Y.}\ \bibnamefont
  {Lu}}, \bibinfo {author} {\bibfnamefont {Y.}~\bibnamefont {Cao}}, \bibinfo
  {author} {\bibfnamefont {C.-Z.}\ \bibnamefont {Peng}},\ and\ \bibinfo
  {author} {\bibfnamefont {J.-W.}\ \bibnamefont {Pan}},\ }\bibfield  {title}
  {\bibinfo {title} {Micius quantum experiments in space},\ }\href
  {https://doi.org/10.1103/RevModPhys.94.035001} {\bibfield  {journal}
  {\bibinfo  {journal} {Rev. Mod. Phys.}\ }\textbf {\bibinfo {volume} {94}},\
  \bibinfo {pages} {035001} (\bibinfo {year} {2022})}\BibitemShut {NoStop}%
\bibitem [{\citenamefont {Li}\ \emph {et~al.}(2025)\citenamefont {Li},
  \citenamefont {Cai}, \citenamefont {Ren},  \emph
  {et~al.}}]{li2024microsatellitebasedrealtimequantumkey}%
  \BibitemOpen
  \bibfield  {author} {\bibinfo {author} {\bibfnamefont {Y.}~\bibnamefont
  {Li}}, \bibinfo {author} {\bibfnamefont {W.-Q.}\ \bibnamefont {Cai}},
  \bibinfo {author} {\bibfnamefont {J.-G.}\ \bibnamefont {Ren}}, , \emph
  {et~al.},\ }\bibfield  {title} {\bibinfo {title} {Microsatellite-based
  real-time quantum key distribution},\ }\href
  {https://doi.org/10.1038/s41586-025-08739-z} {\bibfield  {journal} {\bibinfo
  {journal} {Nature}\ }\textbf {\bibinfo {volume} {640}},\ \bibinfo {pages}
  {47} (\bibinfo {year} {2025})}\BibitemShut {NoStop}%
\bibitem [{\citenamefont {Yin}\ \emph {et~al.}(2012)\citenamefont {Yin},
  \citenamefont {Ren}, \citenamefont {Lu} \emph {et~al.}}]{yin2012quantum}%
  \BibitemOpen
  \bibfield  {author} {\bibinfo {author} {\bibfnamefont {J.}~\bibnamefont
  {Yin}}, \bibinfo {author} {\bibfnamefont {J.-G.}\ \bibnamefont {Ren}},
  \bibinfo {author} {\bibfnamefont {H.}~\bibnamefont {Lu}}, \emph {et~al.},\
  }\bibfield  {title} {\bibinfo {title} {Quantum teleportation and entanglement
  distribution over 100-kilometre free-space channels},\ }\href
  {https://doi.org/https://doi.org/10.1038/nature11332} {\bibfield  {journal}
  {\bibinfo  {journal} {Nature}\ }\textbf {\bibinfo {volume} {488}},\ \bibinfo
  {pages} {185} (\bibinfo {year} {2012})}\BibitemShut {NoStop}%
\bibitem [{\citenamefont {Liu}\ \emph {et~al.}(2024{\natexlab{a}})\citenamefont
  {Liu}, \citenamefont {Luo}, \citenamefont {Yu},\ and\ \citenamefont
  {et~al.}}]{Liu2024}%
  \BibitemOpen
  \bibfield  {author} {\bibinfo {author} {\bibfnamefont {J.~L.}\ \bibnamefont
  {Liu}}, \bibinfo {author} {\bibfnamefont {X.~Y.}\ \bibnamefont {Luo}},
  \bibinfo {author} {\bibfnamefont {Y.}~\bibnamefont {Yu}},\ and\ \bibinfo
  {author} {\bibnamefont {et~al.}},\ }\bibfield  {title} {\bibinfo {title}
  {Creation of memory–memory entanglement in a metropolitan quantum
  network},\ }\href
  {https://doi.org/https://doi.org/10.1038/s41586-024-07308-0} {\bibfield
  {journal} {\bibinfo  {journal} {Nature}\ }\textbf {\bibinfo {volume} {629}},\
  \bibinfo {pages} {579} (\bibinfo {year} {2024}{\natexlab{a}})}\BibitemShut
  {NoStop}%
\bibitem [{\citenamefont {Knaut}\ \emph {et~al.}(2024)\citenamefont {Knaut},
  \citenamefont {Suleymanzade}, \citenamefont {Wei},\ and\ \citenamefont
  {et~al.}}]{Knaut2024}%
  \BibitemOpen
  \bibfield  {author} {\bibinfo {author} {\bibfnamefont {C.~M.}\ \bibnamefont
  {Knaut}}, \bibinfo {author} {\bibfnamefont {A.}~\bibnamefont {Suleymanzade}},
  \bibinfo {author} {\bibfnamefont {Y.~C.}\ \bibnamefont {Wei}},\ and\ \bibinfo
  {author} {\bibnamefont {et~al.}},\ }\bibfield  {title} {\bibinfo {title}
  {Entanglement of nanophotonic quantum memory nodes in a telecom network},\
  }\href {https://doi.org/https://doi.org/10.1038/s41586-024-07252-z}
  {\bibfield  {journal} {\bibinfo  {journal} {Nature}\ }\textbf {\bibinfo
  {volume} {629}},\ \bibinfo {pages} {573} (\bibinfo {year}
  {2024})}\BibitemShut {NoStop}%
\bibitem [{\citenamefont {Bennett}\ \emph {et~al.}(1999)\citenamefont
  {Bennett}, \citenamefont {Shor}, \citenamefont {Smolin},\ and\ \citenamefont
  {Thapliyal}}]{PhysRevLett.83.3081}%
  \BibitemOpen
  \bibfield  {author} {\bibinfo {author} {\bibfnamefont {C.~H.}\ \bibnamefont
  {Bennett}}, \bibinfo {author} {\bibfnamefont {P.~W.}\ \bibnamefont {Shor}},
  \bibinfo {author} {\bibfnamefont {J.~A.}\ \bibnamefont {Smolin}},\ and\
  \bibinfo {author} {\bibfnamefont {A.~V.}\ \bibnamefont {Thapliyal}},\
  }\bibfield  {title} {\bibinfo {title} {Entanglement-assisted classical
  capacity of noisy quantum channels},\ }\href
  {https://doi.org/10.1103/PhysRevLett.83.3081} {\bibfield  {journal} {\bibinfo
   {journal} {Phys. Rev. Lett.}\ }\textbf {\bibinfo {volume} {83}},\ \bibinfo
  {pages} {3081} (\bibinfo {year} {1999})}\BibitemShut {NoStop}%
\bibitem [{\citenamefont {Harrow}\ \emph {et~al.}(2004)\citenamefont {Harrow},
  \citenamefont {Hayden},\ and\ \citenamefont {Leung}}]{PhysRevLett.92.187901}%
  \BibitemOpen
  \bibfield  {author} {\bibinfo {author} {\bibfnamefont {A.}~\bibnamefont
  {Harrow}}, \bibinfo {author} {\bibfnamefont {P.}~\bibnamefont {Hayden}},\
  and\ \bibinfo {author} {\bibfnamefont {D.}~\bibnamefont {Leung}},\ }\bibfield
   {title} {\bibinfo {title} {Superdense coding of quantum states},\ }\href
  {https://doi.org/10.1103/PhysRevLett.92.187901} {\bibfield  {journal}
  {\bibinfo  {journal} {Phys. Rev. Lett.}\ }\textbf {\bibinfo {volume} {92}},\
  \bibinfo {pages} {187901} (\bibinfo {year} {2004})}\BibitemShut {NoStop}%
\bibitem [{\citenamefont {Barreiro}\ \emph {et~al.}(2008)\citenamefont
  {Barreiro}, \citenamefont {Wei},\ and\ \citenamefont {Kwiat}}]{Barreiro2008}%
  \BibitemOpen
  \bibfield  {author} {\bibinfo {author} {\bibfnamefont {J.~T.}\ \bibnamefont
  {Barreiro}}, \bibinfo {author} {\bibfnamefont {T.-C.}\ \bibnamefont {Wei}},\
  and\ \bibinfo {author} {\bibfnamefont {P.~G.}\ \bibnamefont {Kwiat}},\
  }\bibfield  {title} {\bibinfo {title} {Beating the channel capacity limit for
  linear photonic superdense coding},\ }\href
  {https://doi.org/https://doi.org/10.1038/nphys919} {\bibfield  {journal}
  {\bibinfo  {journal} {Nature Physics}\ }\textbf {\bibinfo {volume} {4}},\
  \bibinfo {pages} {282} (\bibinfo {year} {2008})}\BibitemShut {NoStop}%
\bibitem [{\citenamefont {Hu}\ \emph {et~al.}(2018)\citenamefont {Hu},
  \citenamefont {Guo}, \citenamefont {Liu}, \citenamefont {Huang},
  \citenamefont {Li},\ and\ \citenamefont {Guo}}]{Hu2018}%
  \BibitemOpen
  \bibfield  {author} {\bibinfo {author} {\bibfnamefont {X.-M.}\ \bibnamefont
  {Hu}}, \bibinfo {author} {\bibfnamefont {Y.}~\bibnamefont {Guo}}, \bibinfo
  {author} {\bibfnamefont {B.-H.}\ \bibnamefont {Liu}}, \bibinfo {author}
  {\bibfnamefont {Y.-F.}\ \bibnamefont {Huang}}, \bibinfo {author}
  {\bibfnamefont {C.-F.}\ \bibnamefont {Li}},\ and\ \bibinfo {author}
  {\bibfnamefont {G.-C.}\ \bibnamefont {Guo}},\ }\bibfield  {title} {\bibinfo
  {title} {Beating the channel capacity limit for superdense coding with
  entangled ququarts},\ }\href {https://doi.org/10.1126/sciadv.aat9304}
  {\bibfield  {journal} {\bibinfo  {journal} {Science Advances}\ }\textbf
  {\bibinfo {volume} {4}},\ \bibinfo {pages} {eaat9304} (\bibinfo {year}
  {2018})}\BibitemShut {NoStop}%
\bibitem [{\citenamefont {Li}\ \emph {et~al.}(2024)\citenamefont {Li},
  \citenamefont {Xing}, \citenamefont {Qu}, \citenamefont {Xiao}, \citenamefont
  {Fan}, \citenamefont {Zheng}, \citenamefont {Ma}, \citenamefont {Xue},
  \citenamefont {Bharti}, \citenamefont {Koh},\ and\ \citenamefont
  {Xiao}}]{li2024communication}%
  \BibitemOpen
  \bibfield  {author} {\bibinfo {author} {\bibfnamefont {Y.}~\bibnamefont
  {Li}}, \bibinfo {author} {\bibfnamefont {J.}~\bibnamefont {Xing}}, \bibinfo
  {author} {\bibfnamefont {D.}~\bibnamefont {Qu}}, \bibinfo {author}
  {\bibfnamefont {L.}~\bibnamefont {Xiao}}, \bibinfo {author} {\bibfnamefont
  {Z.}~\bibnamefont {Fan}}, \bibinfo {author} {\bibfnamefont {Z.-J.}\
  \bibnamefont {Zheng}}, \bibinfo {author} {\bibfnamefont {H.}~\bibnamefont
  {Ma}}, \bibinfo {author} {\bibfnamefont {P.}~\bibnamefont {Xue}}, \bibinfo
  {author} {\bibfnamefont {K.}~\bibnamefont {Bharti}}, \bibinfo {author}
  {\bibfnamefont {D.~E.}\ \bibnamefont {Koh}},\ and\ \bibinfo {author}
  {\bibfnamefont {Y.}~\bibnamefont {Xiao}},\ }\href@noop {} {\bibinfo {title}
  {Communication with quantum catalysts}} (\bibinfo {year} {2024}),\ \Eprint
  {https://arxiv.org/abs/2406.14395} {arXiv:2406.14395 [quant-ph]} \BibitemShut
  {NoStop}%
\bibitem [{\citenamefont {Knill}\ and\ \citenamefont
  {Laflamme}(1997)}]{PhysRevA.55.900}%
  \BibitemOpen
  \bibfield  {author} {\bibinfo {author} {\bibfnamefont {E.}~\bibnamefont
  {Knill}}\ and\ \bibinfo {author} {\bibfnamefont {R.}~\bibnamefont
  {Laflamme}},\ }\bibfield  {title} {\bibinfo {title} {Theory of quantum
  error-correcting codes},\ }\href {https://doi.org/10.1103/PhysRevA.55.900}
  {\bibfield  {journal} {\bibinfo  {journal} {Phys. Rev. A}\ }\textbf {\bibinfo
  {volume} {55}},\ \bibinfo {pages} {900} (\bibinfo {year} {1997})}\BibitemShut
  {NoStop}%
\bibitem [{\citenamefont {Aoki}\ \emph {et~al.}(2009)\citenamefont {Aoki},
  \citenamefont {Takahashi}, \citenamefont {Kajiya},\ and\ \citenamefont
  {et~al.}}]{Aoki2009}%
  \BibitemOpen
  \bibfield  {author} {\bibinfo {author} {\bibfnamefont {T.}~\bibnamefont
  {Aoki}}, \bibinfo {author} {\bibfnamefont {G.}~\bibnamefont {Takahashi}},
  \bibinfo {author} {\bibfnamefont {T.}~\bibnamefont {Kajiya}},\ and\ \bibinfo
  {author} {\bibnamefont {et~al.}},\ }\bibfield  {title} {\bibinfo {title}
  {Quantum error correction beyond qubits},\ }\href
  {https://doi.org/https://doi.org/10.1038/nphys1309} {\bibfield  {journal}
  {\bibinfo  {journal} {Nature Physics}\ }\textbf {\bibinfo {volume} {5}},\
  \bibinfo {pages} {541} (\bibinfo {year} {2009})}\BibitemShut {NoStop}%
\bibitem [{\citenamefont {Terhal}(2015)}]{RevModPhys.87.307}%
  \BibitemOpen
  \bibfield  {author} {\bibinfo {author} {\bibfnamefont {B.~M.}\ \bibnamefont
  {Terhal}},\ }\bibfield  {title} {\bibinfo {title} {Quantum error correction
  for quantum memories},\ }\href {https://doi.org/10.1103/RevModPhys.87.307}
  {\bibfield  {journal} {\bibinfo  {journal} {Rev. Mod. Phys.}\ }\textbf
  {\bibinfo {volume} {87}},\ \bibinfo {pages} {307} (\bibinfo {year}
  {2015})}\BibitemShut {NoStop}%
\bibitem [{\citenamefont {Krinner}\ \emph {et~al.}(2022)\citenamefont
  {Krinner}, \citenamefont {Lacroix}, \citenamefont {Remm},\ and\ \citenamefont
  {et~al.}}]{Krinner2022}%
  \BibitemOpen
  \bibfield  {author} {\bibinfo {author} {\bibfnamefont {S.}~\bibnamefont
  {Krinner}}, \bibinfo {author} {\bibfnamefont {N.}~\bibnamefont {Lacroix}},
  \bibinfo {author} {\bibfnamefont {A.}~\bibnamefont {Remm}},\ and\ \bibinfo
  {author} {\bibnamefont {et~al.}},\ }\bibfield  {title} {\bibinfo {title}
  {Realizing repeated quantum error correction in a distance-three surface
  code},\ }\href {https://doi.org/https://doi.org/10.1038/s41586-022-04566-8}
  {\bibfield  {journal} {\bibinfo  {journal} {Nature}\ }\textbf {\bibinfo
  {volume} {605}},\ \bibinfo {pages} {669} (\bibinfo {year}
  {2022})}\BibitemShut {NoStop}%
\bibitem [{\citenamefont {Sivak}\ \emph {et~al.}(2023)\citenamefont {Sivak},
  \citenamefont {Eickbusch}, \citenamefont {Royer},\ and\ \citenamefont
  {et~al.}}]{Sivak2023}%
  \BibitemOpen
  \bibfield  {author} {\bibinfo {author} {\bibfnamefont {V.~V.}\ \bibnamefont
  {Sivak}}, \bibinfo {author} {\bibfnamefont {A.}~\bibnamefont {Eickbusch}},
  \bibinfo {author} {\bibfnamefont {B.}~\bibnamefont {Royer}},\ and\ \bibinfo
  {author} {\bibnamefont {et~al.}},\ }\bibfield  {title} {\bibinfo {title}
  {Real-time quantum error correction beyond break-even},\ }\href
  {https://doi.org/https://doi.org/10.1038/s41586-023-05782-6} {\bibfield
  {journal} {\bibinfo  {journal} {Nature}\ }\textbf {\bibinfo {volume} {616}},\
  \bibinfo {pages} {50} (\bibinfo {year} {2023})}\BibitemShut {NoStop}%
\bibitem [{\citenamefont {Liu}\ and\ \citenamefont {Zhou}(2023)}]{Liu2023}%
  \BibitemOpen
  \bibfield  {author} {\bibinfo {author} {\bibfnamefont {Z.-W.}\ \bibnamefont
  {Liu}}\ and\ \bibinfo {author} {\bibfnamefont {S.}~\bibnamefont {Zhou}},\
  }\bibfield  {title} {\bibinfo {title} {Approximate symmetries and quantum
  error correction},\ }\href
  {https://doi.org/https://doi.org/10.1038/s41534-023-00788-4} {\bibfield
  {journal} {\bibinfo  {journal} {npj Quantum Information}\ }\textbf {\bibinfo
  {volume} {9}},\ \bibinfo {pages} {119} (\bibinfo {year} {2023})}\BibitemShut
  {NoStop}%
\bibitem [{\citenamefont {Liu}\ and\ \citenamefont {Zhou}(2024)}]{Liu2024a}%
  \BibitemOpen
  \bibfield  {author} {\bibinfo {author} {\bibfnamefont {Z.-W.}\ \bibnamefont
  {Liu}}\ and\ \bibinfo {author} {\bibfnamefont {S.}~\bibnamefont {Zhou}},\
  }\bibfield  {title} {\bibinfo {title} {A general theory of quantum codes
  connecting quantum computation, complexity and physics},\ }\href
  {https://doi.org/https://doi.org/10.1038/s41567-024-02622-w} {\bibfield
  {journal} {\bibinfo  {journal} {Nature Physics}\ }\textbf {\bibinfo {volume}
  {20}},\ \bibinfo {pages} {1708} (\bibinfo {year} {2024})}\BibitemShut
  {NoStop}%
\bibitem [{\citenamefont {Endo}\ \emph {et~al.}(2018)\citenamefont {Endo},
  \citenamefont {Benjamin},\ and\ \citenamefont {Li}}]{PhysRevX.8.031027}%
  \BibitemOpen
  \bibfield  {author} {\bibinfo {author} {\bibfnamefont {S.}~\bibnamefont
  {Endo}}, \bibinfo {author} {\bibfnamefont {S.~C.}\ \bibnamefont {Benjamin}},\
  and\ \bibinfo {author} {\bibfnamefont {Y.}~\bibnamefont {Li}},\ }\bibfield
  {title} {\bibinfo {title} {Practical quantum error mitigation for near-future
  applications},\ }\href {https://doi.org/10.1103/PhysRevX.8.031027} {\bibfield
   {journal} {\bibinfo  {journal} {Phys. Rev. X}\ }\textbf {\bibinfo {volume}
  {8}},\ \bibinfo {pages} {031027} (\bibinfo {year} {2018})}\BibitemShut
  {NoStop}%
\bibitem [{\citenamefont {Cai}\ \emph {et~al.}(2023)\citenamefont {Cai},
  \citenamefont {Babbush}, \citenamefont {Benjamin}, \citenamefont {Endo},
  \citenamefont {Huggins}, \citenamefont {Li}, \citenamefont {McClean},\ and\
  \citenamefont {O'Brien}}]{RevModPhys.95.045005}%
  \BibitemOpen
  \bibfield  {author} {\bibinfo {author} {\bibfnamefont {Z.}~\bibnamefont
  {Cai}}, \bibinfo {author} {\bibfnamefont {R.}~\bibnamefont {Babbush}},
  \bibinfo {author} {\bibfnamefont {S.~C.}\ \bibnamefont {Benjamin}}, \bibinfo
  {author} {\bibfnamefont {S.}~\bibnamefont {Endo}}, \bibinfo {author}
  {\bibfnamefont {W.~J.}\ \bibnamefont {Huggins}}, \bibinfo {author}
  {\bibfnamefont {Y.}~\bibnamefont {Li}}, \bibinfo {author} {\bibfnamefont
  {J.~R.}\ \bibnamefont {McClean}},\ and\ \bibinfo {author} {\bibfnamefont
  {T.~E.}\ \bibnamefont {O'Brien}},\ }\bibfield  {title} {\bibinfo {title}
  {Quantum error mitigation},\ }\href
  {https://doi.org/10.1103/RevModPhys.95.045005} {\bibfield  {journal}
  {\bibinfo  {journal} {Rev. Mod. Phys.}\ }\textbf {\bibinfo {volume} {95}},\
  \bibinfo {pages} {045005} (\bibinfo {year} {2023})}\BibitemShut {NoStop}%
\bibitem [{\citenamefont {Takagi}\ \emph {et~al.}(2023)\citenamefont {Takagi},
  \citenamefont {Tajima},\ and\ \citenamefont {Gu}}]{PhysRevLett.131.210602}%
  \BibitemOpen
  \bibfield  {author} {\bibinfo {author} {\bibfnamefont {R.}~\bibnamefont
  {Takagi}}, \bibinfo {author} {\bibfnamefont {H.}~\bibnamefont {Tajima}},\
  and\ \bibinfo {author} {\bibfnamefont {M.}~\bibnamefont {Gu}},\ }\bibfield
  {title} {\bibinfo {title} {Universal sampling lower bounds for quantum error
  mitigation},\ }\href {https://doi.org/10.1103/PhysRevLett.131.210602}
  {\bibfield  {journal} {\bibinfo  {journal} {Phys. Rev. Lett.}\ }\textbf
  {\bibinfo {volume} {131}},\ \bibinfo {pages} {210602} (\bibinfo {year}
  {2023})}\BibitemShut {NoStop}%
\bibitem [{\citenamefont {Takagi}\ \emph {et~al.}(2022)\citenamefont {Takagi},
  \citenamefont {Endo}, \citenamefont {Minagawa},\ and\ \citenamefont
  {Gu}}]{Takagi2022}%
  \BibitemOpen
  \bibfield  {author} {\bibinfo {author} {\bibfnamefont {R.}~\bibnamefont
  {Takagi}}, \bibinfo {author} {\bibfnamefont {S.}~\bibnamefont {Endo}},
  \bibinfo {author} {\bibfnamefont {S.}~\bibnamefont {Minagawa}},\ and\
  \bibinfo {author} {\bibfnamefont {M.}~\bibnamefont {Gu}},\ }\bibfield
  {title} {\bibinfo {title} {Fundamental limits of quantum error mitigation},\
  }\href {https://doi.org/https://doi.org/10.1038/s41534-022-00618-z}
  {\bibfield  {journal} {\bibinfo  {journal} {npj Quantum Information}\
  }\textbf {\bibinfo {volume} {8}},\ \bibinfo {pages} {11} (\bibinfo {year}
  {2022})}\BibitemShut {NoStop}%
\bibitem [{\citenamefont {Liu}\ \emph {et~al.}(2024{\natexlab{b}})\citenamefont
  {Liu}, \citenamefont {Xiao},\ and\ \citenamefont
  {Cai}}]{liu2024nonmarkoviannoisesuppressionsimplified}%
  \BibitemOpen
  \bibfield  {author} {\bibinfo {author} {\bibfnamefont {Z.}~\bibnamefont
  {Liu}}, \bibinfo {author} {\bibfnamefont {Y.}~\bibnamefont {Xiao}},\ and\
  \bibinfo {author} {\bibfnamefont {Z.}~\bibnamefont {Cai}},\ }\href@noop {}
  {\bibinfo {title} {Non-markovian noise suppression simplified through channel
  representation}} (\bibinfo {year} {2024}{\natexlab{b}}),\ \Eprint
  {https://arxiv.org/abs/2412.11220} {arXiv:2412.11220 [quant-ph]} \BibitemShut
  {NoStop}%
\bibitem [{\citenamefont {Liao}\ \emph {et~al.}(2025)\citenamefont {Liao},
  \citenamefont {Zhu}, \citenamefont {Chiribella},\ and\ \citenamefont
  {Yang}}]{Liao2025}%
  \BibitemOpen
  \bibfield  {author} {\bibinfo {author} {\bibfnamefont {M.}~\bibnamefont
  {Liao}}, \bibinfo {author} {\bibfnamefont {Y.}~\bibnamefont {Zhu}}, \bibinfo
  {author} {\bibfnamefont {G.}~\bibnamefont {Chiribella}},\ and\ \bibinfo
  {author} {\bibfnamefont {Y.}~\bibnamefont {Yang}},\ }\bibfield  {title}
  {\bibinfo {title} {Noise-agnostic quantum error mitigation with data
  augmented neural models},\ }\href
  {https://doi.org/https://doi.org/10.1038/s41534-025-00960-y} {\bibfield
  {journal} {\bibinfo  {journal} {npj Quantum Information}\ }\textbf {\bibinfo
  {volume} {11}},\ \bibinfo {pages} {8} (\bibinfo {year} {2025})}\BibitemShut
  {NoStop}%
\bibitem [{\citenamefont {Vaswani}\ \emph {et~al.}(2017)\citenamefont
  {Vaswani}, \citenamefont {Shazeer}, \citenamefont {Parmar}, \citenamefont
  {Uszkoreit}, \citenamefont {Jones}, \citenamefont {Gomez}, \citenamefont
  {Kaiser},\ and\ \citenamefont {Polosukhin}}]{NIPS2017_3f5ee243}%
  \BibitemOpen
  \bibfield  {author} {\bibinfo {author} {\bibfnamefont {A.}~\bibnamefont
  {Vaswani}}, \bibinfo {author} {\bibfnamefont {N.}~\bibnamefont {Shazeer}},
  \bibinfo {author} {\bibfnamefont {N.}~\bibnamefont {Parmar}}, \bibinfo
  {author} {\bibfnamefont {J.}~\bibnamefont {Uszkoreit}}, \bibinfo {author}
  {\bibfnamefont {L.}~\bibnamefont {Jones}}, \bibinfo {author} {\bibfnamefont
  {A.~N.}\ \bibnamefont {Gomez}}, \bibinfo {author} {\bibfnamefont {L.~u.}\
  \bibnamefont {Kaiser}},\ and\ \bibinfo {author} {\bibfnamefont
  {I.}~\bibnamefont {Polosukhin}},\ }\bibfield  {title} {\bibinfo {title}
  {Attention is all you need},\ }\href
  {https://proceedings.neurips.cc/paper_files/paper/2017/file/3f5ee243547dee91fbd053c1c4a845aa-Paper.pdf}
  {\bibfield  {journal} {\bibinfo  {journal} {Advances in Neural Information
  Processing Systems}\ }\textbf {\bibinfo {volume} {30}} (\bibinfo {year}
  {2017})}\BibitemShut {NoStop}%
\bibitem [{\citenamefont {Devlin}\ \emph {et~al.}(2019)\citenamefont {Devlin},
  \citenamefont {Chang}, \citenamefont {Lee},\ and\ \citenamefont
  {Toutanova}}]{devlin-etal-2019-bert}%
  \BibitemOpen
  \bibfield  {author} {\bibinfo {author} {\bibfnamefont {J.}~\bibnamefont
  {Devlin}}, \bibinfo {author} {\bibfnamefont {M.-W.}\ \bibnamefont {Chang}},
  \bibinfo {author} {\bibfnamefont {K.}~\bibnamefont {Lee}},\ and\ \bibinfo
  {author} {\bibfnamefont {K.}~\bibnamefont {Toutanova}},\ }\bibfield  {title}
  {\bibinfo {title} {{BERT}: Pre-training of deep bidirectional transformers
  for language understanding},\ }\href {https://doi.org/10.18653/v1/N19-1423}
  {\bibfield  {journal} {\bibinfo  {journal} {Proceedings of the 2019
  Conference of the North {A}merican Chapter of the Association for
  Computational Linguistics: Human Language Technologies, Volume 1}\ ,\
  \bibinfo {pages} {4171}} (\bibinfo {year} {2019})}\BibitemShut {NoStop}%
\bibitem [{\citenamefont {Bello}\ \emph {et~al.}(2023)\citenamefont {Bello},
  \citenamefont {Ng},\ and\ \citenamefont {Leung}}]{bello2023bert}%
  \BibitemOpen
  \bibfield  {author} {\bibinfo {author} {\bibfnamefont {A.}~\bibnamefont
  {Bello}}, \bibinfo {author} {\bibfnamefont {S.-C.}\ \bibnamefont {Ng}},\ and\
  \bibinfo {author} {\bibfnamefont {M.-F.}\ \bibnamefont {Leung}},\ }\bibfield
  {title} {\bibinfo {title} {A bert framework to sentiment analysis of
  tweets},\ }\href {https://doi.org/https://doi.org/10.3390/s23010506}
  {\bibfield  {journal} {\bibinfo  {journal} {Sensors}\ }\textbf {\bibinfo
  {volume} {23}},\ \bibinfo {pages} {506} (\bibinfo {year} {2023})}\BibitemShut
  {NoStop}%
\bibitem [{\citenamefont {Cao}\ and\ \citenamefont
  {Lai}(2020)}]{cao2020bilingual}%
  \BibitemOpen
  \bibfield  {author} {\bibinfo {author} {\bibfnamefont {J.}~\bibnamefont
  {Cao}}\ and\ \bibinfo {author} {\bibfnamefont {C.}~\bibnamefont {Lai}},\
  }\bibfield  {title} {\bibinfo {title} {A bilingual multi-type spam detection
  model based on m-bert},\ }\href
  {https://doi.org/10.1109/GLOBECOM42002.2020.9347970} {\bibfield  {journal}
  {\bibinfo  {journal} {GLOBECOM 2020-2020 IEEE Global Communications
  Conference}\ ,\ \bibinfo {pages} {1}} (\bibinfo {year} {2020})}\BibitemShut
  {NoStop}%
\bibitem [{\citenamefont {Oswald}\ \emph {et~al.}(2022)\citenamefont {Oswald},
  \citenamefont {Simon},\ and\ \citenamefont
  {Bhattacharya}}]{oswald2022spotspam}%
  \BibitemOpen
  \bibfield  {author} {\bibinfo {author} {\bibfnamefont {C.}~\bibnamefont
  {Oswald}}, \bibinfo {author} {\bibfnamefont {S.~E.}\ \bibnamefont {Simon}},\
  and\ \bibinfo {author} {\bibfnamefont {A.}~\bibnamefont {Bhattacharya}},\
  }\bibfield  {title} {\bibinfo {title} {Spotspam: Intention analysis--driven
  sms spam detection using bert embeddings},\ }\href
  {https://doi.org/10.1145/3538491} {\bibfield  {journal} {\bibinfo  {journal}
  {ACM Transactions on the Web (TWEB)}\ }\textbf {\bibinfo {volume} {16}},\
  \bibinfo {pages} {1} (\bibinfo {year} {2022})}\BibitemShut {NoStop}%
\bibitem [{\citenamefont {Hakala}\ and\ \citenamefont
  {Pyysalo}(2019)}]{hakala2019biomedical}%
  \BibitemOpen
  \bibfield  {author} {\bibinfo {author} {\bibfnamefont {K.}~\bibnamefont
  {Hakala}}\ and\ \bibinfo {author} {\bibfnamefont {S.}~\bibnamefont
  {Pyysalo}},\ }\bibfield  {title} {\bibinfo {title} {Biomedical named entity
  recognition with multilingual bert},\ }\href
  {https://doi.org/10.18653/v1/D19-5709} {\bibfield  {journal} {\bibinfo
  {journal} {Proceedings of the 5th workshop on BioNLP open shared tasks}\ ,\
  \bibinfo {pages} {56}} (\bibinfo {year} {2019})}\BibitemShut {NoStop}%
\bibitem [{\citenamefont {Chang}\ \emph {et~al.}(2021)\citenamefont {Chang},
  \citenamefont {Kong}, \citenamefont {Jia},\ and\ \citenamefont
  {Meng}}]{chang2021chinese}%
  \BibitemOpen
  \bibfield  {author} {\bibinfo {author} {\bibfnamefont {Y.}~\bibnamefont
  {Chang}}, \bibinfo {author} {\bibfnamefont {L.}~\bibnamefont {Kong}},
  \bibinfo {author} {\bibfnamefont {K.}~\bibnamefont {Jia}},\ and\ \bibinfo
  {author} {\bibfnamefont {Q.}~\bibnamefont {Meng}},\ }\bibfield  {title}
  {\bibinfo {title} {Chinese named entity recognition method based on bert},\
  }\href {https://doi.org/10.1109/ICDSCA53499.2021.9650256} {\bibfield
  {journal} {\bibinfo  {journal} {2021 IEEE international conference on data
  science and computer application (ICDSCA)}\ ,\ \bibinfo {pages} {294}}
  (\bibinfo {year} {2021})}\BibitemShut {NoStop}%
\bibitem [{\citenamefont {Yang}\ \emph {et~al.}(2019)\citenamefont {Yang},
  \citenamefont {Dai}, \citenamefont {Yang}, \citenamefont {Carbonell},
  \citenamefont {Salakhutdinov},\ and\ \citenamefont
  {Le}}]{10.5555/3454287.3454804}%
  \BibitemOpen
  \bibfield  {author} {\bibinfo {author} {\bibfnamefont {Z.}~\bibnamefont
  {Yang}}, \bibinfo {author} {\bibfnamefont {Z.}~\bibnamefont {Dai}}, \bibinfo
  {author} {\bibfnamefont {Y.}~\bibnamefont {Yang}}, \bibinfo {author}
  {\bibfnamefont {J.}~\bibnamefont {Carbonell}}, \bibinfo {author}
  {\bibfnamefont {R.}~\bibnamefont {Salakhutdinov}},\ and\ \bibinfo {author}
  {\bibfnamefont {Q.~V.}\ \bibnamefont {Le}},\ }\bibfield  {title} {\bibinfo
  {title} {Xlnet: generalized autoregressive pretraining for language
  understanding},\ }\href
  {{https://proceedings.neurips.cc/paper_files/paper/2019/file/dc6a7e655d7e5840e66733e9ee67cc69-Paper.pdf}}
  {\bibfield  {journal} {\bibinfo  {journal} {Proceedings of the 33rd
  International Conference on Neural Information Processing Systems}\ }
  (\bibinfo {year} {2019})}\BibitemShut {NoStop}%
\bibitem [{\citenamefont {Liu}\ \emph {et~al.}(2021)\citenamefont {Liu},
  \citenamefont {Lin}, \citenamefont {Shi},\ and\ \citenamefont
  {Zhao}}]{zhuang-etal-2021-robustly}%
  \BibitemOpen
  \bibfield  {author} {\bibinfo {author} {\bibfnamefont {Z.}~\bibnamefont
  {Liu}}, \bibinfo {author} {\bibfnamefont {W.}~\bibnamefont {Lin}}, \bibinfo
  {author} {\bibfnamefont {Y.}~\bibnamefont {Shi}},\ and\ \bibinfo {author}
  {\bibfnamefont {J.}~\bibnamefont {Zhao}},\ }\bibfield  {title} {\bibinfo
  {title} {A robustly optimized bert pre-training approach with
  post-training},\ }\href {https://doi.org/10.1007/978-3-030-84186-7_31}
  {\bibfield  {journal} {\bibinfo  {journal} {Chinese Computational
  Linguistics}\ }\textbf {\bibinfo {volume} {LNCS 13175}},\ \bibinfo {pages}
  {471} (\bibinfo {year} {2021})}\BibitemShut {NoStop}%
\bibitem [{\citenamefont {Lan}\ \emph {et~al.}(2020)\citenamefont {Lan},
  \citenamefont {Chen}, \citenamefont {Goodman}, \citenamefont {Gimpel},
  \citenamefont {Sharma},\ and\ \citenamefont {Soricut}}]{Lan2019ALBERTAL}%
  \BibitemOpen
  \bibfield  {author} {\bibinfo {author} {\bibfnamefont {Z.}~\bibnamefont
  {Lan}}, \bibinfo {author} {\bibfnamefont {M.}~\bibnamefont {Chen}}, \bibinfo
  {author} {\bibfnamefont {S.}~\bibnamefont {Goodman}}, \bibinfo {author}
  {\bibfnamefont {K.}~\bibnamefont {Gimpel}}, \bibinfo {author} {\bibfnamefont
  {P.}~\bibnamefont {Sharma}},\ and\ \bibinfo {author} {\bibfnamefont
  {R.}~\bibnamefont {Soricut}},\ }\href@noop {} {\bibinfo {title} {Albert: A
  lite bert for self-supervised learning of language representations}}
  (\bibinfo {year} {2020}),\ \Eprint {https://arxiv.org/abs/1909.11942}
  {arXiv:1909.11942 [cs.CL]} \BibitemShut {NoStop}%
\bibitem [{\citenamefont {Hoang}\ \emph {et~al.}(2019)\citenamefont {Hoang},
  \citenamefont {Bihorac},\ and\ \citenamefont {Rouces}}]{hoang2019aspect}%
  \BibitemOpen
  \bibfield  {author} {\bibinfo {author} {\bibfnamefont {M.}~\bibnamefont
  {Hoang}}, \bibinfo {author} {\bibfnamefont {O.~A.}\ \bibnamefont {Bihorac}},\
  and\ \bibinfo {author} {\bibfnamefont {J.}~\bibnamefont {Rouces}},\
  }\bibfield  {title} {\bibinfo {title} {Aspect-based sentiment analysis using
  bert},\ }\href {https://doi.org/https://aclanthology.org/W19-6120/}
  {\bibfield  {journal} {\bibinfo  {journal} {Proceedings of the 22nd nordic
  conference on computational linguistics}\ ,\ \bibinfo {pages} {187}}
  (\bibinfo {year} {2019})}\BibitemShut {NoStop}%
\bibitem [{\citenamefont {Xu}\ \emph {et~al.}(2019)\citenamefont {Xu},
  \citenamefont {Liu}, \citenamefont {Shu},\ and\ \citenamefont
  {Philip}}]{xu2019bert}%
  \BibitemOpen
  \bibfield  {author} {\bibinfo {author} {\bibfnamefont {H.}~\bibnamefont
  {Xu}}, \bibinfo {author} {\bibfnamefont {B.}~\bibnamefont {Liu}}, \bibinfo
  {author} {\bibfnamefont {L.}~\bibnamefont {Shu}},\ and\ \bibinfo {author}
  {\bibfnamefont {S.~Y.}\ \bibnamefont {Philip}},\ }\bibfield  {title}
  {\bibinfo {title} {Bert post-training for review reading comprehension and
  aspect-based sentiment analysis},\ }\href
  {https://doi.org/https://doi.org/10.48550/arXiv.1904.02232} {\bibfield
  {journal} {\bibinfo  {journal} {Proceedings of the 2019 Conference of the
  North American Chapter of the Association for Computational Linguistics:
  Human Language Technologies, Volume 1 (Long and Short Papers)}\ ,\ \bibinfo
  {pages} {2324}} (\bibinfo {year} {2019})}\BibitemShut {NoStop}%
\bibitem [{\citenamefont {Sousa}\ \emph {et~al.}(2019)\citenamefont {Sousa},
  \citenamefont {Sakiyama}, \citenamefont {Rodrigues}, \citenamefont {Moraes},
  \citenamefont {Fernandes},\ and\ \citenamefont {Matsubara}}]{8995193}%
  \BibitemOpen
  \bibfield  {author} {\bibinfo {author} {\bibfnamefont {M.~G.}\ \bibnamefont
  {Sousa}}, \bibinfo {author} {\bibfnamefont {K.}~\bibnamefont {Sakiyama}},
  \bibinfo {author} {\bibfnamefont {L.~d.~S.}\ \bibnamefont {Rodrigues}},
  \bibinfo {author} {\bibfnamefont {P.~H.}\ \bibnamefont {Moraes}}, \bibinfo
  {author} {\bibfnamefont {E.~R.}\ \bibnamefont {Fernandes}},\ and\ \bibinfo
  {author} {\bibfnamefont {E.~T.}\ \bibnamefont {Matsubara}},\ }\bibfield
  {title} {\bibinfo {title} {Bert for stock market sentiment analysis},\ }\href
  {https://doi.org/doi: 10.1109/ICTAI.2019.00231} {\bibfield  {journal}
  {\bibinfo  {journal} {2019 IEEE 31st International Conference on Tools with
  Artificial Intelligence (ICTAI)}\ ,\ \bibinfo {pages} {1597}} (\bibinfo
  {year} {2019})}\BibitemShut {NoStop}%
\bibitem [{\citenamefont {Cao}\ \emph {et~al.}(2021)\citenamefont {Cao},
  \citenamefont {Zhang}, \citenamefont {Chen} \emph {et~al.}}]{Chen2021}%
  \BibitemOpen
  \bibfield  {author} {\bibinfo {author} {\bibfnamefont {Y.}~\bibnamefont
  {Cao}}, \bibinfo {author} {\bibfnamefont {Q.}~\bibnamefont {Zhang}}, \bibinfo
  {author} {\bibfnamefont {T.-Y.}\ \bibnamefont {Chen}}, \emph {et~al.},\
  }\bibfield  {title} {\bibinfo {title} {An integrated space-to-ground quantum
  communication network over 4,600 kilometres},\ }\href
  {https://doi.org/https://doi.org/10.1038/s41586-020-03093-8} {\bibfield
  {journal} {\bibinfo  {journal} {Nature}\ }\textbf {\bibinfo {volume} {589}},\
  \bibinfo {pages} {214} (\bibinfo {year} {2021})}\BibitemShut {NoStop}%
\bibitem [{\citenamefont {Prottasha}\ \emph {et~al.}(2022)\citenamefont
  {Prottasha}, \citenamefont {Sami}, \citenamefont {Kowsher}, \citenamefont
  {Murad}, \citenamefont {Bairagi}, \citenamefont {Masud},\ and\ \citenamefont
  {Baz}}]{prottasha2022transfer}%
  \BibitemOpen
  \bibfield  {author} {\bibinfo {author} {\bibfnamefont {N.~J.}\ \bibnamefont
  {Prottasha}}, \bibinfo {author} {\bibfnamefont {A.~A.}\ \bibnamefont {Sami}},
  \bibinfo {author} {\bibfnamefont {M.}~\bibnamefont {Kowsher}}, \bibinfo
  {author} {\bibfnamefont {S.~A.}\ \bibnamefont {Murad}}, \bibinfo {author}
  {\bibfnamefont {A.~K.}\ \bibnamefont {Bairagi}}, \bibinfo {author}
  {\bibfnamefont {M.}~\bibnamefont {Masud}},\ and\ \bibinfo {author}
  {\bibfnamefont {M.}~\bibnamefont {Baz}},\ }\bibfield  {title} {\bibinfo
  {title} {Transfer learning for sentiment analysis using bert based supervised
  fine-tuning},\ }\href {https://doi.org/https://doi.org/10.3390/s22114157}
  {\bibfield  {journal} {\bibinfo  {journal} {Sensors}\ }\textbf {\bibinfo
  {volume} {22}},\ \bibinfo {pages} {4157} (\bibinfo {year}
  {2022})}\BibitemShut {NoStop}%
\bibitem [{\citenamefont {Zhang}\ \emph {et~al.}(2020)\citenamefont {Zhang},
  \citenamefont {Huang}, \citenamefont {Liu},\ and\ \citenamefont
  {Li}}]{zhang2020spelling}%
  \BibitemOpen
  \bibfield  {author} {\bibinfo {author} {\bibfnamefont {S.}~\bibnamefont
  {Zhang}}, \bibinfo {author} {\bibfnamefont {H.}~\bibnamefont {Huang}},
  \bibinfo {author} {\bibfnamefont {J.}~\bibnamefont {Liu}},\ and\ \bibinfo
  {author} {\bibfnamefont {H.}~\bibnamefont {Li}},\ }\bibfield  {title}
  {\bibinfo {title} {Spelling error correction with soft-masked bert},\ }\href
  {https://doi.org/https://doi.org/10.48550/arXiv.2005.07421} {\bibfield
  {journal} {\bibinfo  {journal} {Proceedings of the 58th Annual Meeting of the
  Association for Computational Linguistics}\ ,\ \bibinfo {pages} {882}}
  (\bibinfo {year} {2020})}\BibitemShut {NoStop}%
\bibitem [{\citenamefont {Tan}\ \emph {et~al.}(2020)\citenamefont {Tan},
  \citenamefont {Chen}, \citenamefont {Li},\ and\ \citenamefont
  {Wang}}]{tan2020spelling}%
  \BibitemOpen
  \bibfield  {author} {\bibinfo {author} {\bibfnamefont {M.}~\bibnamefont
  {Tan}}, \bibinfo {author} {\bibfnamefont {D.}~\bibnamefont {Chen}}, \bibinfo
  {author} {\bibfnamefont {Z.}~\bibnamefont {Li}},\ and\ \bibinfo {author}
  {\bibfnamefont {P.}~\bibnamefont {Wang}},\ }\bibfield  {title} {\bibinfo
  {title} {Spelling error correction with bert based on character-phonetic},\
  }\href {https://doi.org/10.1109/ICCC51575.2020.9345276} {\bibfield  {journal}
  {\bibinfo  {journal} {2020 IEEE 6th International Conference on Computer and
  Communications (ICCC)}\ ,\ \bibinfo {pages} {1146}} (\bibinfo {year}
  {2020})}\BibitemShut {NoStop}%
\bibitem [{\citenamefont {Nguyen}\ \emph {et~al.}(2020)\citenamefont {Nguyen},
  \citenamefont {Jatowt}, \citenamefont {Nguyen}, \citenamefont {Coustaty},\
  and\ \citenamefont {Doucet}}]{nguyen2020neural}%
  \BibitemOpen
  \bibfield  {author} {\bibinfo {author} {\bibfnamefont {T.~T.~H.}\
  \bibnamefont {Nguyen}}, \bibinfo {author} {\bibfnamefont {A.}~\bibnamefont
  {Jatowt}}, \bibinfo {author} {\bibfnamefont {N.-V.}\ \bibnamefont {Nguyen}},
  \bibinfo {author} {\bibfnamefont {M.}~\bibnamefont {Coustaty}},\ and\
  \bibinfo {author} {\bibfnamefont {A.}~\bibnamefont {Doucet}},\ }\bibfield
  {title} {\bibinfo {title} {Neural machine translation with bert for post-ocr
  error detection and correction},\ }\href
  {https://doi.org/10.1145/3383583.3398605} {\bibfield  {journal} {\bibinfo
  {journal} {Proceedings of the ACM/IEEE joint conference on digital libraries
  in 2020}\ ,\ \bibinfo {pages} {333}} (\bibinfo {year} {2020})}\BibitemShut
  {NoStop}%
\bibitem [{\citenamefont {Horodecki}\ \emph {et~al.}(2009)\citenamefont
  {Horodecki}, \citenamefont {Horodecki}, \citenamefont {Horodecki},\ and\
  \citenamefont {Horodecki}}]{RevModPhys.81.865}%
  \BibitemOpen
  \bibfield  {author} {\bibinfo {author} {\bibfnamefont {R.}~\bibnamefont
  {Horodecki}}, \bibinfo {author} {\bibfnamefont {P.}~\bibnamefont
  {Horodecki}}, \bibinfo {author} {\bibfnamefont {M.}~\bibnamefont
  {Horodecki}},\ and\ \bibinfo {author} {\bibfnamefont {K.}~\bibnamefont
  {Horodecki}},\ }\bibfield  {title} {\bibinfo {title} {Quantum entanglement},\
  }\href {https://doi.org/https://doi.org/10.1103/RevModPhys.81.865} {\bibfield
   {journal} {\bibinfo  {journal} {Rev. Mod. Phys.}\ }\textbf {\bibinfo
  {volume} {81}},\ \bibinfo {pages} {865} (\bibinfo {year} {2009})}\BibitemShut
  {NoStop}%
\bibitem [{\citenamefont {Khatri}\ and\ \citenamefont
  {Wilde}(2024)}]{khatri2024principlesquantumcommunicationtheory}%
  \BibitemOpen
  \bibfield  {author} {\bibinfo {author} {\bibfnamefont {S.}~\bibnamefont
  {Khatri}}\ and\ \bibinfo {author} {\bibfnamefont {M.~M.}\ \bibnamefont
  {Wilde}},\ }\href@noop {} {\bibinfo {title} {Principles of quantum
  communication theory: A modern approach}} (\bibinfo {year} {2024}),\ \Eprint
  {https://arxiv.org/abs/2011.04672} {arXiv:2011.04672 [quant-ph]} \BibitemShut
  {NoStop}%
\bibitem [{\citenamefont {Schuck}\ \emph {et~al.}(2006)\citenamefont {Schuck},
  \citenamefont {Huber}, \citenamefont {Kurtsiefer},\ and\ \citenamefont
  {Weinfurter}}]{PhysRevLett.96.190501}%
  \BibitemOpen
  \bibfield  {author} {\bibinfo {author} {\bibfnamefont {C.}~\bibnamefont
  {Schuck}}, \bibinfo {author} {\bibfnamefont {G.}~\bibnamefont {Huber}},
  \bibinfo {author} {\bibfnamefont {C.}~\bibnamefont {Kurtsiefer}},\ and\
  \bibinfo {author} {\bibfnamefont {H.}~\bibnamefont {Weinfurter}},\ }\bibfield
   {title} {\bibinfo {title} {Complete deterministic linear optics bell state
  analysis},\ }\href {https://doi.org/10.1103/PhysRevLett.96.190501} {\bibfield
   {journal} {\bibinfo  {journal} {Phys. Rev. Lett.}\ }\textbf {\bibinfo
  {volume} {96}},\ \bibinfo {pages} {190501} (\bibinfo {year}
  {2006})}\BibitemShut {NoStop}%
\bibitem [{\citenamefont {Hodosh}\ \emph {et~al.}(2013)\citenamefont {Hodosh},
  \citenamefont {Young},\ and\ \citenamefont
  {Hockenmaier}}]{hodosh2013flickr8k}%
  \BibitemOpen
  \bibfield  {author} {\bibinfo {author} {\bibfnamefont {M.}~\bibnamefont
  {Hodosh}}, \bibinfo {author} {\bibfnamefont {P.}~\bibnamefont {Young}},\ and\
  \bibinfo {author} {\bibfnamefont {J.}~\bibnamefont {Hockenmaier}},\
  }\bibfield  {title} {\bibinfo {title} {Framing image description as a ranking
  task: Data, models and evaluation metrics},\ }\href
  {https://doi.org/https://doi.org/10.1613/jair.3994} {\bibfield  {journal}
  {\bibinfo  {journal} {Journal of Artificial Intelligence Research}\ }\textbf
  {\bibinfo {volume} {47}},\ \bibinfo {pages} {853} (\bibinfo {year}
  {2013})}\BibitemShut {NoStop}%
\bibitem [{\citenamefont {Warstadt}\ \emph {et~al.}(2019)\citenamefont
  {Warstadt}, \citenamefont {Singh},\ and\ \citenamefont
  {Bowman}}]{warstadt2019cola}%
  \BibitemOpen
  \bibfield  {author} {\bibinfo {author} {\bibfnamefont {A.}~\bibnamefont
  {Warstadt}}, \bibinfo {author} {\bibfnamefont {A.}~\bibnamefont {Singh}},\
  and\ \bibinfo {author} {\bibfnamefont {S.~R.}\ \bibnamefont {Bowman}},\
  }\bibfield  {title} {\bibinfo {title} {A corpus for modeling rich
  morphological agreement},\ }\href
  {https://doi.org/http://hdl.handle.net/2451/60441} {\bibfield  {journal}
  {\bibinfo  {journal} {Proceedings of the 2019 Conference of the North
  American Chapter of the Association for Computational Linguistics: Human
  Language Technologies, Volume 1 (Long and Short Papers)}\ ,\ \bibinfo {pages}
  {4186}} (\bibinfo {year} {2019})}\BibitemShut {NoStop}%
\bibitem [{\citenamefont {Beltagy}\ \emph {et~al.}(2020)\citenamefont
  {Beltagy}, \citenamefont {Peters},\ and\ \citenamefont
  {Cohan}}]{beltagy2020longformer}%
  \BibitemOpen
  \bibfield  {author} {\bibinfo {author} {\bibfnamefont {I.}~\bibnamefont
  {Beltagy}}, \bibinfo {author} {\bibfnamefont {M.~E.}\ \bibnamefont
  {Peters}},\ and\ \bibinfo {author} {\bibfnamefont {A.}~\bibnamefont
  {Cohan}},\ }\href@noop {} {\bibinfo {title} {Longformer: The long-document
  transformer}} (\bibinfo {year} {2020}),\ \Eprint
  {https://arxiv.org/abs/2004.05150} {arXiv:2004.05150 [cs.CL]} \BibitemShut
  {NoStop}%
\bibitem [{\citenamefont {Zaheer}\ \emph {et~al.}(2021)\citenamefont {Zaheer},
  \citenamefont {Guruganesh}, \citenamefont {Dubey}, \citenamefont {Ainslie},
  \citenamefont {Alberti}, \citenamefont {Ontanon}, \citenamefont {Pham},
  \citenamefont {Ravula}, \citenamefont {Wang}, \citenamefont {Yang},\ and\
  \citenamefont {Ahmed}}]{zaheer2021bigbirdtransformerslonger}%
  \BibitemOpen
  \bibfield  {author} {\bibinfo {author} {\bibfnamefont {M.}~\bibnamefont
  {Zaheer}}, \bibinfo {author} {\bibfnamefont {G.}~\bibnamefont {Guruganesh}},
  \bibinfo {author} {\bibfnamefont {A.}~\bibnamefont {Dubey}}, \bibinfo
  {author} {\bibfnamefont {J.}~\bibnamefont {Ainslie}}, \bibinfo {author}
  {\bibfnamefont {C.}~\bibnamefont {Alberti}}, \bibinfo {author} {\bibfnamefont
  {S.}~\bibnamefont {Ontanon}}, \bibinfo {author} {\bibfnamefont
  {P.}~\bibnamefont {Pham}}, \bibinfo {author} {\bibfnamefont {A.}~\bibnamefont
  {Ravula}}, \bibinfo {author} {\bibfnamefont {Q.}~\bibnamefont {Wang}},
  \bibinfo {author} {\bibfnamefont {L.}~\bibnamefont {Yang}},\ and\ \bibinfo
  {author} {\bibfnamefont {A.}~\bibnamefont {Ahmed}},\ }\href@noop {} {\bibinfo
  {title} {Big bird: Transformers for longer sequences}} (\bibinfo {year}
  {2021}),\ \Eprint {https://arxiv.org/abs/2007.14062} {arXiv:2007.14062
  [cs.CL]} \BibitemShut {NoStop}%
\bibitem [{\citenamefont {Chatterjee}\ \emph {et~al.}(2023)\citenamefont
  {Chatterjee}, \citenamefont {Phalak},\ and\ \citenamefont
  {Ghosh}}]{10313604}%
  \BibitemOpen
  \bibfield  {author} {\bibinfo {author} {\bibfnamefont {A.}~\bibnamefont
  {Chatterjee}}, \bibinfo {author} {\bibfnamefont {K.}~\bibnamefont {Phalak}},\
  and\ \bibinfo {author} {\bibfnamefont {S.}~\bibnamefont {Ghosh}},\ }\bibfield
   {title} {\bibinfo {title} {Quantum error correction for dummies},\ }\href
  {https://doi.org/10.1109/QCE57702.2023.00017} {\bibfield  {journal} {\bibinfo
   {journal} {2023 IEEE International Conference on Quantum Computing and
  Engineering (QCE)}\ }\textbf {\bibinfo {volume} {01}},\ \bibinfo {pages} {70}
  (\bibinfo {year} {2023})}\BibitemShut {NoStop}%
\bibitem [{\citenamefont {Carrasquilla}(2020)}]{carrasquilla2020machine}%
  \BibitemOpen
  \bibfield  {author} {\bibinfo {author} {\bibfnamefont {J.}~\bibnamefont
  {Carrasquilla}},\ }\bibfield  {title} {\bibinfo {title} {Machine learning for
  quantum matter},\ }\href
  {https://doi.org/https://doi.org/10.1080/23746149.2020.1797528} {\bibfield
  {journal} {\bibinfo  {journal} {Advances in Physics: X}\ }\textbf {\bibinfo
  {volume} {5}},\ \bibinfo {pages} {1797528} (\bibinfo {year}
  {2020})}\BibitemShut {NoStop}%
\bibitem [{\citenamefont {Melko}\ \emph {et~al.}(2019)\citenamefont {Melko},
  \citenamefont {Carleo}, \citenamefont {Carrasquilla},\ and\ \citenamefont
  {Cirac}}]{melko2019restricted}%
  \BibitemOpen
  \bibfield  {author} {\bibinfo {author} {\bibfnamefont {R.~G.}\ \bibnamefont
  {Melko}}, \bibinfo {author} {\bibfnamefont {G.}~\bibnamefont {Carleo}},
  \bibinfo {author} {\bibfnamefont {J.}~\bibnamefont {Carrasquilla}},\ and\
  \bibinfo {author} {\bibfnamefont {J.~I.}\ \bibnamefont {Cirac}},\ }\bibfield
  {title} {\bibinfo {title} {Restricted boltzmann machines in quantum
  physics},\ }\href {https://doi.org/https://doi.org/10.1038/s41567-019-0545-1}
  {\bibfield  {journal} {\bibinfo  {journal} {Nature Physics}\ }\textbf
  {\bibinfo {volume} {15}},\ \bibinfo {pages} {887} (\bibinfo {year}
  {2019})}\BibitemShut {NoStop}%
\bibitem [{\citenamefont {Carleo}\ and\ \citenamefont
  {Troyer}(2017)}]{carleo2017solving}%
  \BibitemOpen
  \bibfield  {author} {\bibinfo {author} {\bibfnamefont {G.}~\bibnamefont
  {Carleo}}\ and\ \bibinfo {author} {\bibfnamefont {M.}~\bibnamefont
  {Troyer}},\ }\bibfield  {title} {\bibinfo {title} {Solving the quantum
  many-body problem with artificial neural networks},\ }\href
  {https://doi.org/10.1126/science.aag2302} {\bibfield  {journal} {\bibinfo
  {journal} {Science}\ }\textbf {\bibinfo {volume} {355}},\ \bibinfo {pages}
  {602} (\bibinfo {year} {2017})}\BibitemShut {NoStop}%
\bibitem [{\citenamefont {Carrasquilla}\ and\ \citenamefont
  {Melko}(2017)}]{carrasquilla2017machine}%
  \BibitemOpen
  \bibfield  {author} {\bibinfo {author} {\bibfnamefont {J.}~\bibnamefont
  {Carrasquilla}}\ and\ \bibinfo {author} {\bibfnamefont {R.~G.}\ \bibnamefont
  {Melko}},\ }\bibfield  {title} {\bibinfo {title} {Machine learning phases of
  matter},\ }\href
  {https://doi.org/https://doi.org/10.1080/23746149.2020.1797528} {\bibfield
  {journal} {\bibinfo  {journal} {Nature Physics}\ }\textbf {\bibinfo {volume}
  {13}},\ \bibinfo {pages} {431} (\bibinfo {year} {2017})}\BibitemShut
  {NoStop}%
\bibitem [{\citenamefont {Walln\"ofer}\ \emph {et~al.}(2020)\citenamefont
  {Walln\"ofer}, \citenamefont {Melnikov}, \citenamefont {D\"ur},\ and\
  \citenamefont {Briegel}}]{PRXQuantum.1.010301}%
  \BibitemOpen
  \bibfield  {author} {\bibinfo {author} {\bibfnamefont {J.}~\bibnamefont
  {Walln\"ofer}}, \bibinfo {author} {\bibfnamefont {A.~A.}\ \bibnamefont
  {Melnikov}}, \bibinfo {author} {\bibfnamefont {W.}~\bibnamefont {D\"ur}},\
  and\ \bibinfo {author} {\bibfnamefont {H.~J.}\ \bibnamefont {Briegel}},\
  }\bibfield  {title} {\bibinfo {title} {Machine learning for long-distance
  quantum communication},\ }\href {https://doi.org/10.1103/PRXQuantum.1.010301}
  {\bibfield  {journal} {\bibinfo  {journal} {PRX Quantum}\ }\textbf {\bibinfo
  {volume} {1}},\ \bibinfo {pages} {010301} (\bibinfo {year}
  {2020})}\BibitemShut {NoStop}%
\bibitem [{\citenamefont {Biamonte}\ \emph {et~al.}(2017)\citenamefont
  {Biamonte}, \citenamefont {Wittek}, \citenamefont {Pancotti}, \citenamefont
  {Rebentrost}, \citenamefont {Wiebe},\ and\ \citenamefont
  {Lloyd}}]{biamonte2017quantum}%
  \BibitemOpen
  \bibfield  {author} {\bibinfo {author} {\bibfnamefont {J.}~\bibnamefont
  {Biamonte}}, \bibinfo {author} {\bibfnamefont {P.}~\bibnamefont {Wittek}},
  \bibinfo {author} {\bibfnamefont {N.}~\bibnamefont {Pancotti}}, \bibinfo
  {author} {\bibfnamefont {P.}~\bibnamefont {Rebentrost}}, \bibinfo {author}
  {\bibfnamefont {N.}~\bibnamefont {Wiebe}},\ and\ \bibinfo {author}
  {\bibfnamefont {S.}~\bibnamefont {Lloyd}},\ }\bibfield  {title} {\bibinfo
  {title} {Quantum machine learning},\ }\href
  {https://doi.org/https://doi.org/10.1038/nature23474} {\bibfield  {journal}
  {\bibinfo  {journal} {Nature}\ }\textbf {\bibinfo {volume} {549}},\ \bibinfo
  {pages} {195} (\bibinfo {year} {2017})}\BibitemShut {NoStop}%
\bibitem [{\citenamefont {Dunjko}\ and\ \citenamefont
  {Briegel}(2018)}]{dunjko2018machine}%
  \BibitemOpen
  \bibfield  {author} {\bibinfo {author} {\bibfnamefont {V.}~\bibnamefont
  {Dunjko}}\ and\ \bibinfo {author} {\bibfnamefont {H.~J.}\ \bibnamefont
  {Briegel}},\ }\bibfield  {title} {\bibinfo {title} {Machine learning \&
  artificial intelligence in the quantum domain: a review of recent progress},\
  }\href {https://doi.org/10.1088/1361-6633/aab406} {\bibfield  {journal}
  {\bibinfo  {journal} {Reports on Progress in Physics}\ }\textbf {\bibinfo
  {volume} {81}},\ \bibinfo {pages} {074001} (\bibinfo {year}
  {2018})}\BibitemShut {NoStop}%
\bibitem [{\citenamefont {Guo}\ \emph {et~al.}(2024)\citenamefont {Guo},
  \citenamefont {Yu}, \citenamefont {Choi}, \citenamefont {Agrawal},
  \citenamefont {Nakaji}, \citenamefont {Aspuru-Guzik},\ and\ \citenamefont
  {Rebentrost}}]{guo2024quantum}%
  \BibitemOpen
  \bibfield  {author} {\bibinfo {author} {\bibfnamefont {N.}~\bibnamefont
  {Guo}}, \bibinfo {author} {\bibfnamefont {Z.}~\bibnamefont {Yu}}, \bibinfo
  {author} {\bibfnamefont {M.}~\bibnamefont {Choi}}, \bibinfo {author}
  {\bibfnamefont {A.}~\bibnamefont {Agrawal}}, \bibinfo {author} {\bibfnamefont
  {K.}~\bibnamefont {Nakaji}}, \bibinfo {author} {\bibfnamefont
  {A.}~\bibnamefont {Aspuru-Guzik}},\ and\ \bibinfo {author} {\bibfnamefont
  {P.}~\bibnamefont {Rebentrost}},\ }\href@noop {} {\bibinfo {title} {Quantum
  linear algebra is all you need for transformer architectures}} (\bibinfo
  {year} {2024}),\ \Eprint {https://arxiv.org/abs/2402.16714} {arXiv:2402.16714
  [quant-ph]} \BibitemShut {NoStop}%
\bibitem [{\citenamefont {Yu}\ \emph {et~al.}(2024)\citenamefont {Yu},
  \citenamefont {Chen}, \citenamefont {Jiao}, \citenamefont {Li}, \citenamefont
  {Lu}, \citenamefont {Wang},\ and\ \citenamefont {Yang}}]{yu2023provable}%
  \BibitemOpen
  \bibfield  {author} {\bibinfo {author} {\bibfnamefont {Z.}~\bibnamefont
  {Yu}}, \bibinfo {author} {\bibfnamefont {Q.}~\bibnamefont {Chen}}, \bibinfo
  {author} {\bibfnamefont {Y.}~\bibnamefont {Jiao}}, \bibinfo {author}
  {\bibfnamefont {Y.}~\bibnamefont {Li}}, \bibinfo {author} {\bibfnamefont
  {X.}~\bibnamefont {Lu}}, \bibinfo {author} {\bibfnamefont {X.}~\bibnamefont
  {Wang}},\ and\ \bibinfo {author} {\bibfnamefont {J.~Z.}\ \bibnamefont
  {Yang}},\ }\href@noop {} {\bibinfo {title} {Non-asymptotic approximation
  error bounds of parameterized quantum circuits}} (\bibinfo {year} {2024}),\
  \Eprint {https://arxiv.org/abs/2310.07528} {arXiv:2310.07528 [quant-ph]}
  \BibitemShut {NoStop}%
\bibitem [{\citenamefont {Cherrat}\ \emph {et~al.}(2024)\citenamefont
  {Cherrat}, \citenamefont {Kerenidis}, \citenamefont {Mathur}, \citenamefont
  {Landman}, \citenamefont {Strahm},\ and\ \citenamefont {Li}}]{Cherrat_2024}%
  \BibitemOpen
  \bibfield  {author} {\bibinfo {author} {\bibfnamefont {E.~A.}\ \bibnamefont
  {Cherrat}}, \bibinfo {author} {\bibfnamefont {I.}~\bibnamefont {Kerenidis}},
  \bibinfo {author} {\bibfnamefont {N.}~\bibnamefont {Mathur}}, \bibinfo
  {author} {\bibfnamefont {J.}~\bibnamefont {Landman}}, \bibinfo {author}
  {\bibfnamefont {M.}~\bibnamefont {Strahm}},\ and\ \bibinfo {author}
  {\bibfnamefont {Y.~Y.}\ \bibnamefont {Li}},\ }\bibfield  {title} {\bibinfo
  {title} {Quantum vision transformers},\ }\href
  {https://doi.org/https://doi.org/10.22331/q-2024-02-22-1265} {\bibfield
  {journal} {\bibinfo  {journal} {Quantum}\ }\textbf {\bibinfo {volume} {8}},\
  \bibinfo {pages} {1265} (\bibinfo {year} {2024})}\BibitemShut {NoStop}%
\bibitem [{\citenamefont {Tian}\ \emph {et~al.}(2023)\citenamefont {Tian},
  \citenamefont {Sun}, \citenamefont {Du}, \citenamefont {Zhao}, \citenamefont
  {Liu}, \citenamefont {Zhang}, \citenamefont {Yi}, \citenamefont {Huang},
  \citenamefont {Wang}, \citenamefont {Wu}, \citenamefont {Hsieh},
  \citenamefont {Liu}, \citenamefont {Yang},\ and\ \citenamefont
  {Tao}}]{tian2023recent}%
  \BibitemOpen
  \bibfield  {author} {\bibinfo {author} {\bibfnamefont {J.}~\bibnamefont
  {Tian}}, \bibinfo {author} {\bibfnamefont {X.}~\bibnamefont {Sun}}, \bibinfo
  {author} {\bibfnamefont {Y.}~\bibnamefont {Du}}, \bibinfo {author}
  {\bibfnamefont {S.}~\bibnamefont {Zhao}}, \bibinfo {author} {\bibfnamefont
  {Q.}~\bibnamefont {Liu}}, \bibinfo {author} {\bibfnamefont {K.}~\bibnamefont
  {Zhang}}, \bibinfo {author} {\bibfnamefont {W.}~\bibnamefont {Yi}}, \bibinfo
  {author} {\bibfnamefont {W.}~\bibnamefont {Huang}}, \bibinfo {author}
  {\bibfnamefont {C.}~\bibnamefont {Wang}}, \bibinfo {author} {\bibfnamefont
  {X.}~\bibnamefont {Wu}}, \bibinfo {author} {\bibfnamefont {M.-H.}\
  \bibnamefont {Hsieh}}, \bibinfo {author} {\bibfnamefont {T.}~\bibnamefont
  {Liu}}, \bibinfo {author} {\bibfnamefont {W.}~\bibnamefont {Yang}},\ and\
  \bibinfo {author} {\bibfnamefont {D.}~\bibnamefont {Tao}},\ }\bibfield
  {title} {\bibinfo {title} {Recent advances for quantum neural networks in
  generative learning},\ }\href {https://doi.org/10.1109/TPAMI.2023.3272029}
  {\bibfield  {journal} {\bibinfo  {journal} {IEEE Transactions on Pattern
  Analysis and Machine Intelligence}\ }\textbf {\bibinfo {volume} {45}},\
  \bibinfo {pages} {12321} (\bibinfo {year} {2023})}\BibitemShut {NoStop}%
\bibitem [{\citenamefont {Wang}\ \emph {et~al.}(2023)\citenamefont {Wang},
  \citenamefont {Liu}, \citenamefont {Liu}, \citenamefont {Luo}, \citenamefont
  {Du},\ and\ \citenamefont {Tao}}]{wang2023symmetric}%
  \BibitemOpen
  \bibfield  {author} {\bibinfo {author} {\bibfnamefont {X.}~\bibnamefont
  {Wang}}, \bibinfo {author} {\bibfnamefont {J.}~\bibnamefont {Liu}}, \bibinfo
  {author} {\bibfnamefont {T.}~\bibnamefont {Liu}}, \bibinfo {author}
  {\bibfnamefont {Y.}~\bibnamefont {Luo}}, \bibinfo {author} {\bibfnamefont
  {Y.}~\bibnamefont {Du}},\ and\ \bibinfo {author} {\bibfnamefont
  {D.}~\bibnamefont {Tao}},\ }\href@noop {} {\bibinfo {title} {Symmetric
  pruning in quantum neural networks}} (\bibinfo {year} {2023}),\ \Eprint
  {https://arxiv.org/abs/2208.14057} {arXiv:2208.14057 [quant-ph]} \BibitemShut
  {NoStop}%
\bibitem [{\citenamefont {Slussarenko}\ and\ \citenamefont
  {Pryde}(2019)}]{10.1063/1.5115814}%
  \BibitemOpen
  \bibfield  {author} {\bibinfo {author} {\bibfnamefont {S.}~\bibnamefont
  {Slussarenko}}\ and\ \bibinfo {author} {\bibfnamefont {G.~J.}\ \bibnamefont
  {Pryde}},\ }\bibfield  {title} {\bibinfo {title} {Photonic quantum
  information processing: A concise review},\ }\href
  {https://doi.org/10.1063/1.5115814} {\bibfield  {journal} {\bibinfo
  {journal} {Applied Physics Reviews}\ }\textbf {\bibinfo {volume} {6}},\
  \bibinfo {pages} {041303} (\bibinfo {year} {2019})}\BibitemShut {NoStop}%
\bibitem [{\citenamefont {Fabre}\ and\ \citenamefont
  {Treps}(2020)}]{RevModPhys.92.035005}%
  \BibitemOpen
  \bibfield  {author} {\bibinfo {author} {\bibfnamefont {C.}~\bibnamefont
  {Fabre}}\ and\ \bibinfo {author} {\bibfnamefont {N.}~\bibnamefont {Treps}},\
  }\bibfield  {title} {\bibinfo {title} {Modes and states in quantum optics},\
  }\href {https://doi.org/10.1103/RevModPhys.92.035005} {\bibfield  {journal}
  {\bibinfo  {journal} {Rev. Mod. Phys.}\ }\textbf {\bibinfo {volume} {92}},\
  \bibinfo {pages} {035005} (\bibinfo {year} {2020})}\BibitemShut {NoStop}%
\bibitem [{\citenamefont {Yuan}\ \emph {et~al.}(2023)\citenamefont {Yuan},
  \citenamefont {Xiao}, \citenamefont {Hou}, \citenamefont {Fei}, \citenamefont
  {Gour}, \citenamefont {Xiang}, \citenamefont {Li},\ and\ \citenamefont
  {Guo}}]{Yuan2023}%
  \BibitemOpen
  \bibfield  {author} {\bibinfo {author} {\bibfnamefont {Y.}~\bibnamefont
  {Yuan}}, \bibinfo {author} {\bibfnamefont {Y.}~\bibnamefont {Xiao}}, \bibinfo
  {author} {\bibfnamefont {Z.}~\bibnamefont {Hou}}, \bibinfo {author}
  {\bibfnamefont {S.-M.}\ \bibnamefont {Fei}}, \bibinfo {author} {\bibfnamefont
  {G.}~\bibnamefont {Gour}}, \bibinfo {author} {\bibfnamefont {G.-Y.}\
  \bibnamefont {Xiang}}, \bibinfo {author} {\bibfnamefont {C.-F.}\ \bibnamefont
  {Li}},\ and\ \bibinfo {author} {\bibfnamefont {G.-C.}\ \bibnamefont {Guo}},\
  }\bibfield  {title} {\bibinfo {title} {Strong majorization uncertainty
  relations and experimental verifications},\ }\href
  {https://doi.org/10.1038/s41534-023-00736-2} {\bibfield  {journal} {\bibinfo
  {journal} {npj Quantum Information}\ }\textbf {\bibinfo {volume} {9}},\
  \bibinfo {pages} {65} (\bibinfo {year} {2023})}\BibitemShut {NoStop}%
\bibitem [{\citenamefont {Rivas}\ and\ \citenamefont
  {Huelga}(2012)}]{Rivas_2012}%
  \BibitemOpen
  \bibfield  {author} {\bibinfo {author} {\bibfnamefont {A.}~\bibnamefont
  {Rivas}}\ and\ \bibinfo {author} {\bibfnamefont {S.~F.}\ \bibnamefont
  {Huelga}},\ }\href
  {https://doi.org/https://doi.org/10.1007/978-3-642-23354-8} {\emph {\bibinfo
  {title} {Open Quantum Systems: An Introduction}}}\ (\bibinfo  {publisher}
  {Springer},\ \bibinfo {year} {2012})\BibitemShut {NoStop}%
\bibitem [{\citenamefont {Pollock}\ \emph {et~al.}(2018)\citenamefont
  {Pollock}, \citenamefont {Rodr\'{\i}guez-Rosario}, \citenamefont
  {Frauenheim}, \citenamefont {Paternostro},\ and\ \citenamefont
  {Modi}}]{PhysRevA.97.012127}%
  \BibitemOpen
  \bibfield  {author} {\bibinfo {author} {\bibfnamefont {F.~A.}\ \bibnamefont
  {Pollock}}, \bibinfo {author} {\bibfnamefont {C.}~\bibnamefont
  {Rodr\'{\i}guez-Rosario}}, \bibinfo {author} {\bibfnamefont {T.}~\bibnamefont
  {Frauenheim}}, \bibinfo {author} {\bibfnamefont {M.}~\bibnamefont
  {Paternostro}},\ and\ \bibinfo {author} {\bibfnamefont {K.}~\bibnamefont
  {Modi}},\ }\bibfield  {title} {\bibinfo {title} {Non-markovian quantum
  processes: Complete framework and efficient characterization},\ }\href
  {https://doi.org/10.1103/PhysRevA.97.012127} {\bibfield  {journal} {\bibinfo
  {journal} {Phys. Rev. A}\ }\textbf {\bibinfo {volume} {97}},\ \bibinfo
  {pages} {012127} (\bibinfo {year} {2018})}\BibitemShut {NoStop}%
\bibitem [{\citenamefont {Lidar}(2020)}]{lidar2020lecturenotestheoryopen}%
  \BibitemOpen
  \bibfield  {author} {\bibinfo {author} {\bibfnamefont {D.~A.}\ \bibnamefont
  {Lidar}},\ }\href@noop {} {\bibinfo {title} {Lecture notes on the theory of
  open quantum systems}} (\bibinfo {year} {2020}),\ \Eprint
  {https://arxiv.org/abs/1902.00967} {arXiv:1902.00967 [quant-ph]} \BibitemShut
  {NoStop}%
\bibitem [{\citenamefont {Milz}\ and\ \citenamefont
  {Modi}(2021)}]{PRXQuantum.2.030201}%
  \BibitemOpen
  \bibfield  {author} {\bibinfo {author} {\bibfnamefont {S.}~\bibnamefont
  {Milz}}\ and\ \bibinfo {author} {\bibfnamefont {K.}~\bibnamefont {Modi}},\
  }\bibfield  {title} {\bibinfo {title} {Quantum stochastic processes and
  quantum non-markovian phenomena},\ }\href
  {https://doi.org/10.1103/PRXQuantum.2.030201} {\bibfield  {journal} {\bibinfo
   {journal} {PRX Quantum}\ }\textbf {\bibinfo {volume} {2}},\ \bibinfo {pages}
  {030201} (\bibinfo {year} {2021})}\BibitemShut {NoStop}%
\bibitem [{\citenamefont {Xiao}\ \emph {et~al.}(2021)\citenamefont {Xiao},
  \citenamefont {Sengupta}, \citenamefont {Yang},\ and\ \citenamefont
  {Gour}}]{PhysRevResearch.3.023077}%
  \BibitemOpen
  \bibfield  {author} {\bibinfo {author} {\bibfnamefont {Y.}~\bibnamefont
  {Xiao}}, \bibinfo {author} {\bibfnamefont {K.}~\bibnamefont {Sengupta}},
  \bibinfo {author} {\bibfnamefont {S.}~\bibnamefont {Yang}},\ and\ \bibinfo
  {author} {\bibfnamefont {G.}~\bibnamefont {Gour}},\ }\bibfield  {title}
  {\bibinfo {title} {Uncertainty principle of quantum processes},\ }\href
  {https://doi.org/10.1103/PhysRevResearch.3.023077} {\bibfield  {journal}
  {\bibinfo  {journal} {Phys. Rev. Res.}\ }\textbf {\bibinfo {volume} {3}},\
  \bibinfo {pages} {023077} (\bibinfo {year} {2021})}\BibitemShut {NoStop}%
\bibitem [{\citenamefont {Xiao}\ \emph {et~al.}(2023)\citenamefont {Xiao},
  \citenamefont {Yang}, \citenamefont {Wang}, \citenamefont {Liu},\ and\
  \citenamefont {Gu}}]{PhysRevLett.130.240201}%
  \BibitemOpen
  \bibfield  {author} {\bibinfo {author} {\bibfnamefont {Y.}~\bibnamefont
  {Xiao}}, \bibinfo {author} {\bibfnamefont {Y.}~\bibnamefont {Yang}}, \bibinfo
  {author} {\bibfnamefont {X.}~\bibnamefont {Wang}}, \bibinfo {author}
  {\bibfnamefont {Q.}~\bibnamefont {Liu}},\ and\ \bibinfo {author}
  {\bibfnamefont {M.}~\bibnamefont {Gu}},\ }\bibfield  {title} {\bibinfo
  {title} {Quantum uncertainty principles for measurements with
  interventions},\ }\href {https://doi.org/10.1103/PhysRevLett.130.240201}
  {\bibfield  {journal} {\bibinfo  {journal} {Phys. Rev. Lett.}\ }\textbf
  {\bibinfo {volume} {130}},\ \bibinfo {pages} {240201} (\bibinfo {year}
  {2023})}\BibitemShut {NoStop}%
\bibitem [{\citenamefont {Taranto}\ \emph {et~al.}(2025)\citenamefont
  {Taranto}, \citenamefont {Milz}, \citenamefont {Murao}, \citenamefont
  {Quintino},\ and\ \citenamefont
  {Modi}}]{taranto2025higherorderquantumoperations}%
  \BibitemOpen
  \bibfield  {author} {\bibinfo {author} {\bibfnamefont {P.}~\bibnamefont
  {Taranto}}, \bibinfo {author} {\bibfnamefont {S.}~\bibnamefont {Milz}},
  \bibinfo {author} {\bibfnamefont {M.}~\bibnamefont {Murao}}, \bibinfo
  {author} {\bibfnamefont {M.~T.}\ \bibnamefont {Quintino}},\ and\ \bibinfo
  {author} {\bibfnamefont {K.}~\bibnamefont {Modi}},\ }\href@noop {} {\bibinfo
  {title} {Higher-order quantum operations}} (\bibinfo {year} {2025}),\ \Eprint
  {https://arxiv.org/abs/2503.09693} {arXiv:2503.09693 [quant-ph]} \BibitemShut
  {NoStop}%
\bibitem [{\citenamefont {Gottesman}(1998)}]{PhysRevA.57.127}%
  \BibitemOpen
  \bibfield  {author} {\bibinfo {author} {\bibfnamefont {D.}~\bibnamefont
  {Gottesman}},\ }\bibfield  {title} {\bibinfo {title} {Theory of
  fault-tolerant quantum computation},\ }\href
  {https://doi.org/10.1103/PhysRevA.57.127} {\bibfield  {journal} {\bibinfo
  {journal} {Phys. Rev. A}\ }\textbf {\bibinfo {volume} {57}},\ \bibinfo
  {pages} {127} (\bibinfo {year} {1998})}\BibitemShut {NoStop}%
\bibitem [{\citenamefont {Lidar}\ and\ \citenamefont
  {Brun}(2013)}]{QECLidar2013}%
  \BibitemOpen
  \bibinfo {editor} {\bibfnamefont {D.~A.}\ \bibnamefont {Lidar}}\ and\
  \bibinfo {editor} {\bibfnamefont {T.~A.}\ \bibnamefont {Brun}},\ eds.,\ \href
  {https://doi.org/10.1142/9789813237230_0010} {\emph {\bibinfo {title}
  {Quantum Error Correction}}}\ (\bibinfo  {publisher} {Cambridge University
  Press},\ \bibinfo {address} {Cambridge},\ \bibinfo {year} {2013})\BibitemShut
  {NoStop}%
\bibitem [{\citenamefont {Gottesman}(2024)}]{gottesman2024surviving}%
  \BibitemOpen
  \bibfield  {author} {\bibinfo {author} {\bibfnamefont {D.}~\bibnamefont
  {Gottesman}},\ }\bibfield  {title} {\bibinfo {title} {Surviving as a quantum
  computer in a classical world},\ }\href
  {https://www.cs.umd.edu/class/spring2024/cmsc858G/QECCbook-2024-ch1-15.pdf}
  {\bibfield  {journal} {\bibinfo  {journal} {Textbook manuscript preprint}\ }
  (\bibinfo {year} {2024})}\BibitemShut {NoStop}%
\bibitem [{\citenamefont {Ross}\ and\ \citenamefont
  {Doll{\'a}r}(2017)}]{ross2017focal}%
  \BibitemOpen
  \bibfield  {author} {\bibinfo {author} {\bibfnamefont {T.-Y.}\ \bibnamefont
  {Ross}}\ and\ \bibinfo {author} {\bibfnamefont {G.}~\bibnamefont
  {Doll{\'a}r}},\ }\bibfield  {title} {\bibinfo {title} {Focal loss for dense
  object detection},\ }\href
  {https://doi.org/https://doi.org/10.48550/arXiv.1708.02002} {\bibfield
  {journal} {\bibinfo  {journal} {proceedings of the IEEE conference on
  computer vision and pattern recognition}\ ,\ \bibinfo {pages} {2980}}
  (\bibinfo {year} {2017})}\BibitemShut {NoStop}%
\bibitem [{\citenamefont {Quijano}\ \emph {et~al.}(2021)\citenamefont
  {Quijano}, \citenamefont {Nguyen},\ and\ \citenamefont
  {Ordonez}}]{Quijano2021GridSH}%
  \BibitemOpen
  \bibfield  {author} {\bibinfo {author} {\bibfnamefont {A.~J.}\ \bibnamefont
  {Quijano}}, \bibinfo {author} {\bibfnamefont {S.}~\bibnamefont {Nguyen}},\
  and\ \bibinfo {author} {\bibfnamefont {J.}~\bibnamefont {Ordonez}},\
  }\href@noop {} {\bibinfo {title} {Grid search hyperparameter benchmarking of
  bert, albert, and longformer on duorc}} (\bibinfo {year} {2021}),\ \Eprint
  {https://arxiv.org/abs/2101.06326} {arXiv:2101.06326 [cs.CL]} \BibitemShut
  {NoStop}%
\end{thebibliography}%


\appendix
\renewcommand{\thesection}{\Alph{section}}
\renewcommand{\thesubsection}{\Alph{section}\arabic{subsection}}


\section{Noise Models}\label{sec:appendix-A}

In a closed quantum system, the evolution of the quantum state is reversible and governed by unitary operations. In contrast, the dynamics of an open quantum system~\cite{Rivas_2012,PhysRevA.97.012127,lidar2020lecturenotestheoryopen,PRXQuantum.2.030201,PhysRevResearch.3.023077,PhysRevLett.130.240201,taranto2025higherorderquantumoperations} -- where interactions with an external environment are considered -- are described by completely positive and trace-preserving (CPTP) linear maps, known as quantum channels. Mathematically, any quantum channel $\mE$ can be represented in terms of Kraus operators as follows
\begin{align}\label{s-eq:noisychannel}
    \mE(\cdot)
    =
    \sum_{i}K_i \cdot K_i^{\dagger},
\end{align}
where the operators satisfy the completeness condition, namely 
\begin{align}
    \sum_i K_i^\dagger K_i = \1,
\end{align}
with $^{\dagger}$ denoting the Hermitian adjoint, and $\1$ representing the identity operator. The Kraus decompositions for various qubit noise models~\cite{PhysRevA.57.127,QECLidar2013,gottesman2024surviving} are listed in Table~\ref{table:noise}.

\begin{table}[h]
\centering
\begin{tabular}{cc}
\toprule
\textbf{Noise Models} & \textbf{Kraus Operators} \\
\hline\hline\addlinespace
\makecell{Bit Flip } & \makecell{
$K_0 = \sqrt{1 - \lambda} I, \quad K_1 = \sqrt{\lambda} X$} \\
\addlinespace\hline\addlinespace
\makecell{Phase Flip } & \makecell{
$K_0 = \sqrt{1 - \lambda} I, \quad K_1 = \sqrt{\lambda} Z$} \\
\addlinespace\hline\addlinespace
\makecell{Depolarizing } & \makecell{
$K_0 = \sqrt{1 - 3\lambda/4} I, \quad K_1 = \sqrt{\lambda/4} X$, \\
$K_2 = \sqrt{\lambda/4} Y, \quad K_3 = \sqrt{\lambda/4} Z$} \\
\addlinespace\hline\addlinespace
\makecell{Amplitude Damping } & \makecell{
$K_0 = \begin{pmatrix} 1 & 0 \\ 0 & \sqrt{1 - \lambda} \end{pmatrix}, K_1 = \begin{pmatrix} 0 & \sqrt{\lambda} \\ 0 & 0 \end{pmatrix}$} \\
\addlinespace\hline\addlinespace
\makecell{Qudit Bit Flip ($d=4$)} & \makecell{$K_0 = \sqrt{1-\lambda} I, \quad K_1 = \sqrt{\lambda/3}X(1)$, \\
$K_2 = \sqrt{\lambda/3} X(2), \quad K_3 = \sqrt{\lambda/3} X(3)$.} \\
\addlinespace\bottomrule
\end{tabular}
\caption{\textbf{Noise Models.} 
The first four rows display widely used noise models for qubits, while the final row outlines the bit-flip error for qudits with $d=4$. The parameter $\lambda$ denotes the noise strength, quantifying the intensity of noise influencing the communication process.
}
\label{table:noise}
\end{table}

The bit-flip error, a fundamental concept in classical information theory, arises when bits are flipped with a probability $\lambda$ during transmission. Beyond the qubit case, we extend our analysis to the qudit scenario with $d=4$, where we conduct numerical experiments to explore the impact of bit-flip errors in subsequent sections. The Kraus operators characterizing this channel are provided in Table~\ref{table:noise}, as discussed in~\cite{khatri2024principlesquantumcommunicationtheory}.

\begin{figure*}[t]
\centering
\begin{algorithm}[H]\label{algorithm:1}
\caption{Correction-Evaluation Network Joint Training}
\label{alg:bert_eval_training}
\KwIn{Dataset $D = \{(\mT_n, \mT^w_n)\}_{n=1}^{N}$: target texts $\mT_n$ and WLRM corrected texts $\mT^w_n$, Initial parameters $\theta$, epochs $K$, learning rate $\gamma$}
\KwOut{Optimized parameters $\theta^*$}

Initialize $\theta \gets \text{Get Pre-training-parameters}()$ \tcp*[r]{Weight Initialization}

\For{$i \gets 1$ \KwTo $N_{\text{epochs}}$}{
    \For{$\text{epoch} \gets 1$ \KwTo $K$}{
        Shuffle $D$ \tcp*[r]{Stochastic Sampling}
        \For{each $(\mT, \mT^w) \in D$}{
            $P_{\text{correction}} \gets \text{Correction\_Network}(\mT^w; \theta)$ \tcp*[r]{Correction Output}
            $P_{\text{evaluation}} \gets \text{Evaluation\_Network}(\mT, \mT^w; \theta)$ \tcp*[r]{Evaluation Output}
            
            $\mL_c \gets \text{CrossEntropy}(P_{\text{correction}}, \mT)$ \;
            $\mL_e \gets \text{Evaluation}(P_{\text{correction}}, P_{\text{evaluation}})$ \;
            
            $\mL \gets \alpha \mL_c + \beta \mL_e$ \tcp*[r]{Combined loss}
            $\theta \gets \theta - \gamma \nabla_{\theta} \mL$ \tcp*[r]{Gradient Descent Update}
        }
        $\theta^* \gets \text{EMA}(\theta)$ \tcp*[r]{Exponential Moving Average}
    }
}
\Return $\theta^*$
\end{algorithm}
\end{figure*}


\section{Correction Network}\label{sec:appendix-B}

The correction network consists of $12$ stacked identical blocks, each comprising a multi-head self-attention layer followed by a feedforward network. Every word $w_i^{w}$ in the input $\mT^{w}$ is transformed into an embedding vector, combined with positional embeddings to encode word order, and marked with special tokens [CLS] at the start and [SEP] at the end. The multi-head self-attention mechanism, described in~\cite{NIPS2017_3f5ee243}, allows the model to capture global dependencies between words in the sequence, irrespective of their distance. 
For an input token embedding matrix  $X \in \mathbb{R}^{n \times d}$ (where  $n$  is the sequence length and $d$ is the embedding dimension), the model deploys  $h$ parallel attention heads. Each head processes the input via three learnable projections 
\begin{align}
\mathrm{head}_i = \mathrm{softmax}\left( \frac{(X W_i^Q)(X W_i^K)^\top}{\sqrt{d/h}} \right) X W_i^V.
\end{align}
The matrices $ W_i^Q, W_i^K, W_i^V \in \mathbb{R}^{d \times d/h}$, with $d/h$ ensuring dimensional consistency across the attention heads. This scaled dot-product attention computes the similarity between queries and keys, scaled by  $\sqrt{d/h}$ to prevent large values that could destabilize gradients.
The attention outputs from each head are then concatenated and projected back to the original embedding dimension
\begin{align}
\mathrm{MultiHead}(X) = \mathrm{Concat}(\mathrm{head}_1, \dots, \mathrm{head}_h) W^O,
\end{align}
where $W^O \in \mathbb{R}^{d \times d}$. The matrix $W^O$ is the output projection matrix that linearly combines the results from all heads, preserving the original embedding space.

Following the self-attention mechanism, the output is combined with the input through a residual connection, followed by layer normalization
\begin{align}
x_{\text{out}} = \mathrm{LayerNorm}(x_{\text{in}} + \mathrm{MultiHead}(x_{\text{in}})),
\end{align}
where $ \mathrm{MultiHead}(x_{\text{in}})$ denotes the output of the multi-head self-attention layer. 
Subsequently, the output is passed through a position-wise feed-forward network (FFN), which operates independently on each position. The FFN is defined as
\begin{align}
\mathrm{FFN}(x) = \max(0, x W_1 + b_1) W_2 + b_2.
\end{align}
Here, $W_1$ and $W_2$ represent learnable weight matrices, and $b_1$, $ b_2$ are the corresponding biases. The FFN output is then added to its input via another residual connection, followed by layer normalization
\begin{align}
x_{\text{out}} = \mathrm{LayerNorm}(x_{\text{out}} + \mathrm{FFN}(x_{\text{out}})).
\end{align}

\begin{figure*}[t]
\centering
\includegraphics[width=1\textwidth]{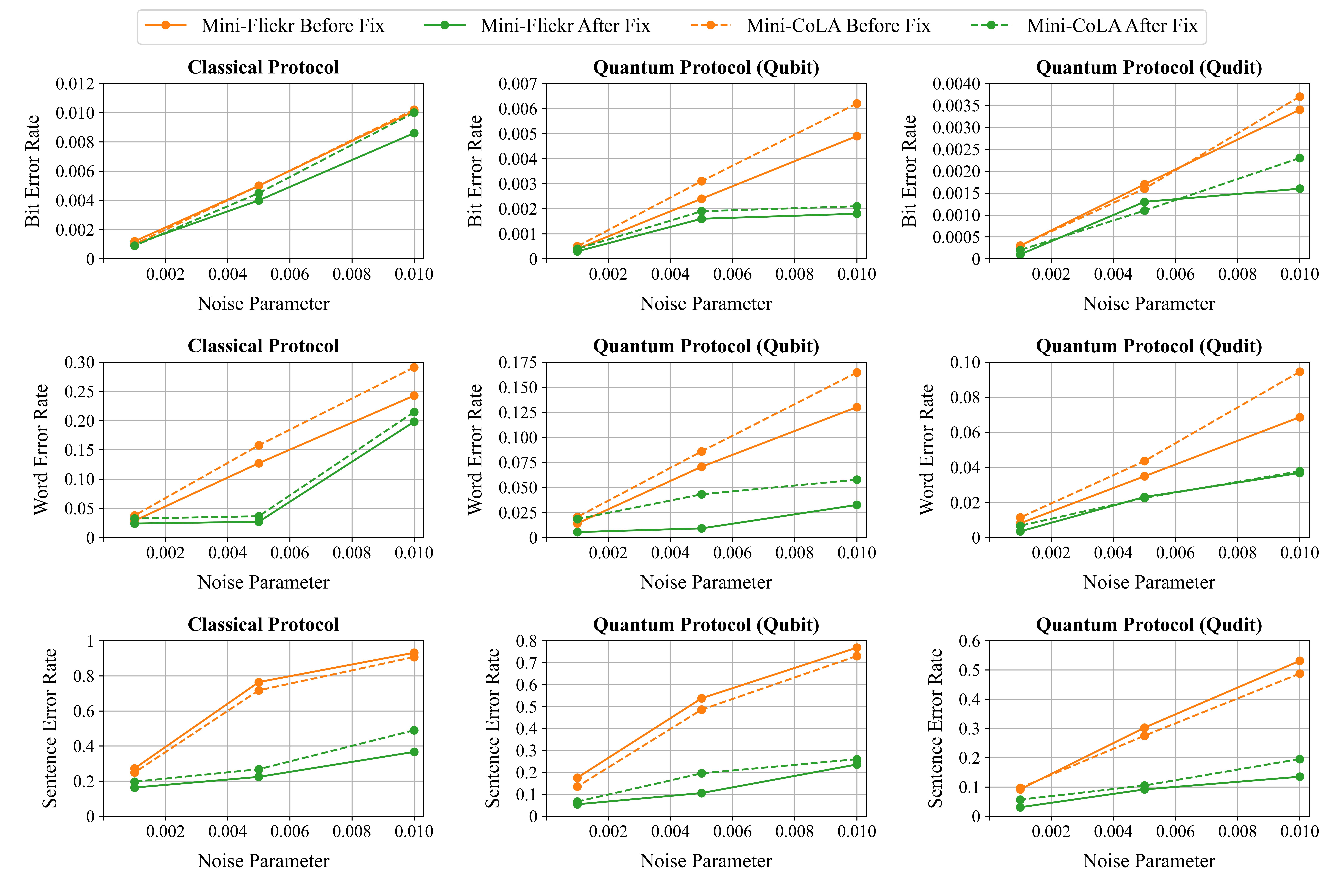}
\captionsetup{justification=raggedright, singlelinecheck=false}
\caption{{\bf Performance Analysis of Language-Model-Assisted Communication.} 
We assess the performance of text transmission across classical bit communication, quantum qubit communication, and quantum qudit communication ($d=4$) under varying bit-flip noise parameters, using two distinct datasets. The orange line represents the error rate prior to the application of the PQC-BERT module, while the green line shows the error rate following its application. Solid lines denote results from the Mini-Flickr dataset, and dashed lines represent results from the Mini-CoLA dataset.
}
\label{fig:classical-quantum2}
\end{figure*}


\section{Training Setup}\label{sec:appendix-C}

The learning phase is focused on the SLRM component. The training dataset consists of input-output pairs $(\mT^{w}, \mT)$, where $\mT^{w}$ denotes the texts corrected by the WLRM, and $\mT$ represents the original texts that the sender, Alice, intends to transmit to the receiver, Bob.
The learning objective is composed of two key components: correction and evaluation. To address these, we formulate two separate loss functions -- one for the correction network and another for the evaluation network. The loss function $\mL_c$ for the correction network is defined as
\begin{align}\label{eq:lc}
    \mL_c:= -\sum_{i=1}^n\log{P(w^c_i|\mT^w)},
\end{align}
where $P(w^c_i|\mT^w)$ denotes the probability of selecting the word $w^c_i$ given the input $\mT^w$. The loss function $\mL_e$ for the evaluation network is based on Focal Loss~\cite{ross2017focal}. Given that incorrect words in $\mT^{q}$ represent a relatively small portion of our dataset, we employ a modified Focal Loss to address the challenge of class imbalance. Using the word-level repaired text $\mT^{w}$, we create a boolean array $q = [q_1, \ldots, q_n]$, where $q_i:= \mathbf{1}\{w^w_i = w_i\}$, with $\mathbf{1}\{\cdot\}$ representing the indicator function, which takes the value 1 when the condition holds, and 0 otherwise.  
Meanwhile, the modified focal loss function is defined as
\begin{align}\label{eq:le}
   \mL_e :=  -\sum_{\mathrm{i}=1}^n\alpha(1-f_i)^\gamma\log(f_i + \epsilon),
\end{align}
where $f_i = c_i$ when $q_i = 1$ (indicating that the word should be replaced), and $f_i = 1 - c_i$ otherwise (indicating that the word should not be replaced). The parameters $\alpha$, $\gamma$, and $\epsilon$ are employed to fine-tune the Focal Loss function, thereby enhancing its effectiveness in handling the imbalanced dataset. The training loss function is expressed as a linear combination of the terms in Eq.~\ref{eq:lc} and Eq.~\ref{eq:le}.
\begin{align}
    \mL := \theta \cdot \mL_c +
    (1 - \theta) \cdot \mL_e,\quad \theta \in [0,1].
\end{align}
The remaining hyperparameters and learning rates were established based on prior work~\cite{Quijano2021GridSH}. We selected the Adam optimizer and implemented slight adjustments to enhance its performance.

Our numerical experiments are based on processed datasets, Mini-Flickr and Mini-CoLA, which are masked with simulated quantum noise. Following extracted $12459$ corrected English sentences from Flickr-8k and $9078$ from CoLA, textual features are mapped to quantum state representations through ASCII encoding. 
Simulated quantum noise is introduced by applying probabilistic noise to randomly selected qubits, with the noise probability determined through random number generation. The data is then split into training, validation, and test sets in an 80:10:10 ratio.  The detailed steps of training is presented in
Algorithm~\ref{algorithm:1}.


\section{Evaluation Metrics}\label{sec:appendix-E}

We classify the test data into positive and negative samples, where positive samples require correction and negative samples do not. When the model successfully corrects a positive sample, it is labeled as a True Positive (TP). If the model fails to correct a positive sample, it is labeled as a False Negative (FN). Correctly processing a negative sample is considered a True Negative (TN), whereas incorrectly processing a negative sample is labeled as a False Positive (FP).

The model was evaluated using the following metrics: {\it Accuracy}, {\it Recall}, {\it Precision}, and {\it F1-score}, defined as follows

\begin{align}
    \text{Accuracy} &= \frac{\text{TP} + \text{TN}}{\text{TP} + \text{FP} + \text{TN} + \text{FN}},\label{eq:metrics_1}\\
    \text{Recall} &= \frac{\text{TP}}{\text{TP} + \text{FN}}\label{eq:metrics_2},\\
    \text{Precision} &= \frac{\text{TP}}{\text{TP} + \text{FP}}\label{eq:metrics_3},\\
    \text{F1-score} &= 2 \times \frac{\text{Precision} \times \text{Recall}}{\text{Precision} + \text{Recall}}.\label{eq:metrics_4}
\end{align}
Accuracy (see Eq.~\ref{eq:metrics_1}) quantifies the proportion of correctly classified sentences. Recall (see Eq.~\ref{eq:metrics_2}) evaluates the model's capacity to identify positive samples, while Precision (see Eq.~\ref{eq:metrics_3}) represents the ratio of true positives among the predicted positives, serving as an indicator of reliability. The F1-score (see Eq.~\ref{eq:metrics_4}) integrates Precision and Recall, offering a balanced measure of the model's overall performance.

\begin{figure*}[ht]
\centering
\includegraphics[width=0.91\textwidth]{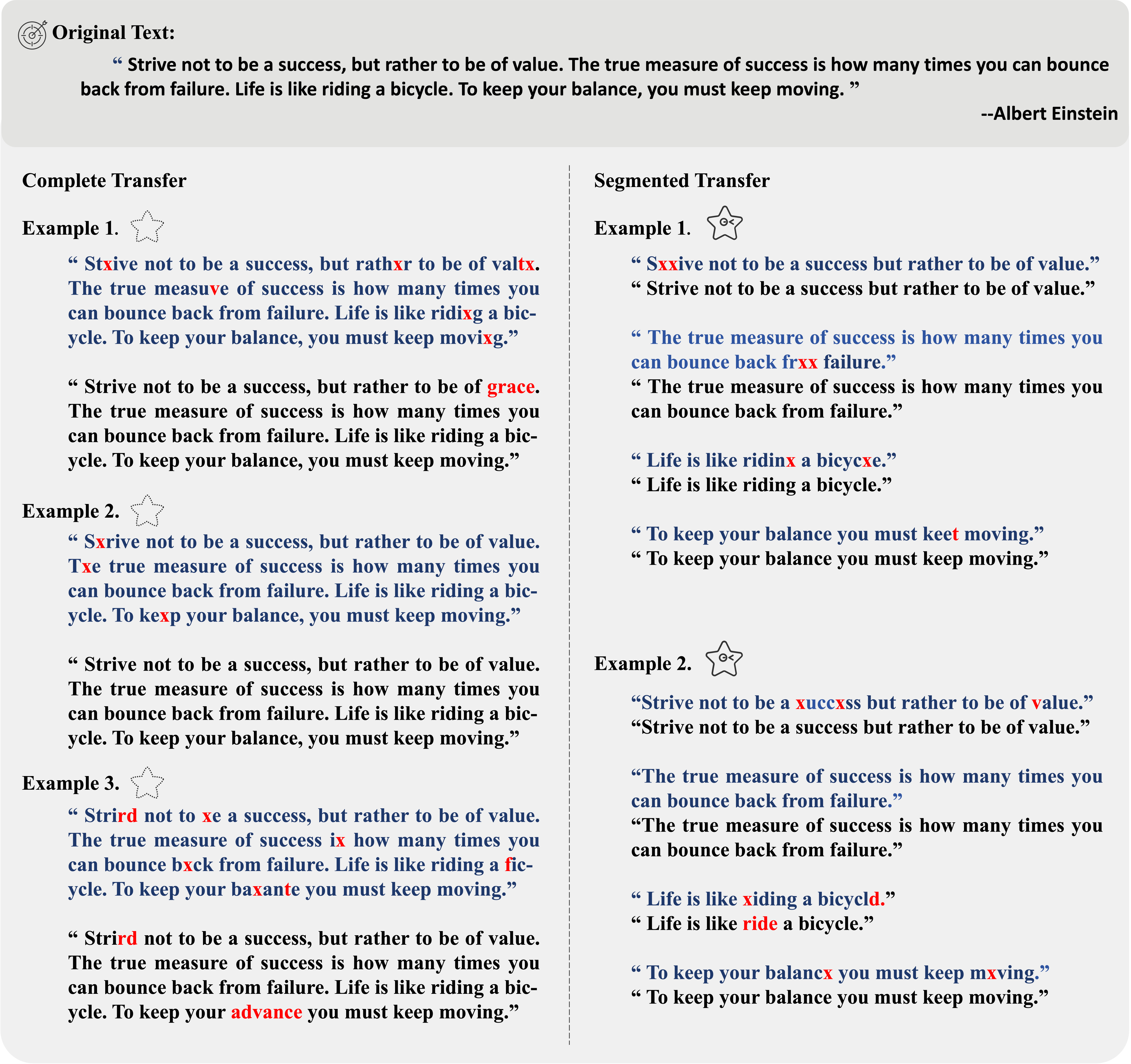}
\captionsetup{justification=raggedright, singlelinecheck=false}
\caption{{\bf Complete Transfer vs Segmented Transfer.} 
Blue text corresponds to transmission via standard superdense coding, where the information is decoded without the application of PQC-BERT, while black text represents transmission utilizing PQC-BERT, with errors distinctly marked in red.
}
\label{fig:text-transmission-long}
\end{figure*}


\section{Performance Analysis}\label{sec:appendix-F}

We present a systematic comparison between classical and quantum communication protocols across a varying range of noise strengths ($0.001$ to $0.01$), using three metrics -- bit, word, and sentence error rates -- to capture performance at increasingly semantic levels. As shown in Fig.~\ref{fig:classical-quantum2}, the numerical results reveal a clear advantage of using PQC-BERT in noisy environments, demonstrating its superior noise tolerance and its effectiveness in preserving the semantic structure of transmitted information.

A key observation from our numerical experiments is that quantum protocols -- both qubit- and qudit-based -- consistently outperform their classical counterparts under identical noise conditions. Even when subjected to the same bit-flip noise model with fixed noise strength, quantum-enhanced communication achieves markedly lower sentence error rates across all tested regimes. For instance, with a noise parameter of $0.001$, a language-model-assisted classical protocol yields a sentence error rate of $13.61\%$, whereas its quantum counterpart using qubits reduces this to $5.42\%$. Notably, further gains are observed when the information is encoded in higher-dimensional systems: replacing qubits with qudits of dimension $d=4$ lowers the error rate to $3.05\%$. These results highlight the role of entanglement and high-dimensional encoding in enhancing the semantic fidelity of communication, suggesting a fundamental advantage of quantum protocols in preserving structured information under noise.


\section{Segmented Transfer}\label{sec:appendix-G}

Written communication can proceed through two contrasting modes: one relying on extended sentences enriched with subordinate clauses to convey layered meaning, and the other favoring sequences of short, declarative statements that unfold information incrementally. These stylistic choices impose different cognitive demands and shape how efficiently a reader processes content. Analogously, in the transmission of textual data, one may adopt a {\it Complete Transfer} strategy -- delivering the entire message in a single block -- or a {\it Segmented Transfer} approach, wherein the message is partitioned into concise units of 5 to 16 words (see Fig.~\ref{fig:length}) and conveyed sequentially. Here, we investigate the performance of segmented transmission, focusing on its potential advantages in fidelity and interpretability under realistic noise conditions.

As a case study, we investigate the transmission of the following quote by Albert Einstein: ·``{\it Strive not to be a success, but rather to be of value. The true measure of success is how many times you can bounce back from failure. Life is like riding a bicycle. To keep your balance, you must keep moving}.'' The message is encoded into a $656$ qubits sequence and sent via superdense coding (see Fig.~\hyperref[fig:SDC-BERT]{1(b)}) over a bit-flip noisy channel with an error probability of $0.01$. Fig.~\ref{fig:text-transmission-long} presents a comparative analysis of two transmission strategies. The left panel shows the outcome under complete transfer, where the full sequence is transmitted as a single block. Across multiple trials, we observe that when noise simultaneously affects several qubits corresponding to letters within the same word, PQC-BERT becomes less effective, resulting in loss of semantic integrity. In contrast, the right panel illustrates the segmented transfer, in which the message is partitioned into smaller units before transmission. This segmentation leads to substantially improved resilience, with localized errors more effectively corrected and the overall intelligibility of the message better preserved. We also note that occurrences of the character ``x'' in the decoded text serve as generic error indicators: they do not represent corrupted instances of the letter ``x'' {\it per se}, but rather stand in for any non-alphabetic character introduced by noise (e.g., ``$\&$'', ``$\$$''), standardized here for interpretability.


\end{document}